\documentclass[%
twocolumn, 
superscriptaddress,
showpacs,
preprintnumbers,
showkeys,
amsmath,
amssymb,
aps,
prb,
floatfix]{revtex4-2}

\usepackage{graphicx} 
\usepackage{amsmath}
\usepackage{mathtools}
\usepackage{xcolor}
\usepackage{graphicx}
\usepackage{bm}
\usepackage[colorlinks=true, citecolor=blue, urlcolor=blue, linkcolor=blue]{hyperref}

\usepackage[normalem]{ulem}
\usepackage{soul}
\usepackage{siunitx}
\usepackage{multirow}
\usepackage{enumitem}
\usepackage{diagbox} 
\usepackage{booktabs}

\newcommand{\CCOC}{Ca$_2$CuO$_2$Cl$_2$}
\newcommand{\NaCCOC}{Na$_x$Ca$_{2-x}$CuO$_2$Cl$_2$}

\newcommand{\LSCO}{La$_{2-x}$Sr$_x$CuO$_4$}

\newcommand\cre{\hat{c}^\dagger}
\newcommand\anh{\hat{c}^{\phantom{\dagger}}}

\newcommand{\SF}{Spin-Flip}
\newcommand{\NSF}{Non-Spin-Flip}

\DeclarePairedDelimiter\bra{\langle}{\rvert}
\DeclarePairedDelimiter\ket{\lvert}{\rangle}

\begin{document}

\title{Paramagnon softening upon doping in oxychloride cuprate superconductors}

\author{Chafic Fawaz}
\thanks{These two authors contributed equally.}
\email{chafic.fawaz@kit.edu}
\altaffiliation{Current address: Institute for Quantum Materials and Technologies, Karlsruhe Institute of Technology, D-76021 Karlsruhe, Germany}
\affiliation{Univ. Grenoble Alpes, CNRS, Grenoble INP, Institut Néel, F-38000 Grenoble, France}

\author{L\'eo Gaspard}
\thanks{These two authors contributed equally.}
\affiliation{IMPMC, Sorbonne Universit\'e, CNRS, MNHN, 4 place Jussieu, F-75252 Paris, France}

\author{Laura Chaix}
\affiliation{Univ. Grenoble Alpes, CNRS, Grenoble INP, Institut Néel, F-38000 Grenoble, France}

\author{Luca Laveder}
\affiliation{Univ. Grenoble Alpes, CNRS, Grenoble INP, Institut Néel, F-38000 Grenoble, France}
\altaffiliation{Current address: IBM}

\author{Benjamin Bacq-Labreuil}
\affiliation{Universit\'e de Strasbourg, CNRS, Institut de Physique et Chimie des Mat\'eriaux de Strasbourg, UMR 7504, F-67000 Strasbourg, France}
\affiliation{D\'epartement de physique, Regroupement qu\'eb\'ecois sur les mat\'eriaux de pointe $\&$ Institut quantique Universit\'e de Sherbrooke, 2500 Boul. Universit\'e, Sherbrooke, Qu\'ebec J1K2R1, Canada}

\author{Victor Por\'ee}
\altaffiliation{Present address: Univ Rennes, CNRS, Institut des Sciences Chimiques de Rennes-UMR6226, 35042 Rennes, France}
\affiliation{Synchrotron SOLEIL, L’Orme des Merisiers, Saint-Aubin, 91192 Gif-sur-Yvette Cedex, France}

\author{Flora Yakhou-Harris}
\affiliation{European Synchrotron Radiation Facility (ESRF), B.P. 220, F-38043 Grenoble Cedex, France}

\author{Kurt Kummer}
\affiliation{European Synchrotron Radiation Facility (ESRF), B.P. 220, F-38043 Grenoble Cedex, France}

\author{Nicholas B. Brookes}
\affiliation{European Synchrotron Radiation Facility (ESRF), B.P. 220, F-38043 Grenoble Cedex, France}

\author{Alessandro Nicolaou}
\affiliation{Synchrotron SOLEIL, L’Orme des Merisiers, Saint-Aubin, 91192 Gif-sur-Yvette Cedex, France}

\author{Hajime Yamamoto}
\altaffiliation[Permanent address: ]{Institute of Multidisciplinary Research for Advanced Materials (IMRAM), Tohoku Univ. Katahira 2-1-1, Aoba-ku, Sendai 980-8577, Japan}
\affiliation{Laboratory for Materials and Structures, Tokyo Institute of Technology, 4259 Nagatsuta, Midori-ku, Yokohama, 226-8503, Japan}

\author{Yuichi Okazaki}
\affiliation{Department of Materials Science, Graduate School of Engineering, Osaka Prefecture University-1-1 Gakuen-cho, Naka-ku, Sakai, Osaka 599-8570, Japan}

\author{Ikuya Yamada}
\affiliation{Department of Materials Science, Graduate School of Engineering, Osaka Prefecture University-1-1 Gakuen-cho, Naka-ku, Sakai, Osaka 599-8570, Japan}

\author{Masaki Azuma}
\affiliation{Materials and Structures Laboratory, Institute of Integrated Research, Institute of Science Tokyo, 4259 Nagatsuta, Midori-ku, Yokohama 226-8503, Japan}

\author{David Santos-Cottin}
\affiliation{Department of Physics, University of Fribourg, 1700 Fribourg, Switzerland}

\author{Marie-Aude M\'easson}
\affiliation{Univ. Grenoble Alpes, CNRS, Grenoble INP, Institut Néel, F-38000 Grenoble, France}

\author{Benjamin Lenz}
\email{benjamin.lenz@upmc.fr}
\affiliation{IMPMC, Sorbonne Universit\'e, CNRS, MNHN, 4 place Jussieu, F-75252 Paris, France}

\author{Matteo d'Astuto}
\email{matteo.dastuto@neel.cnrs.fr}
\affiliation{Univ. Grenoble Alpes, CNRS, Grenoble INP, Institut Néel, F-38000 Grenoble, France}

\date{\today}

\begin{abstract}
Understanding how magnetic excitations and exchange interactions evolve with doping is crucial for deciphering the physics that shapes the phase diagram of high-temperature cuprate superconductors. 
Here, we clearly determine the behavior of the paramagnon dispersion in the oxychloride cuprate Na$_{x}$Ca$_{2-x}$CuO$_2$Cl$_2$ up to the highest doping level. 
We find that the paramagnon bandwidth, defined by the energy at the $X$ zone boundary $(0.5,0,0)$, remains unchanged within experimental uncertainty, even though it exhibits significant broadening. In contrast, along the nodal direction $\Gamma-M$ (\textit{i.e.}, from $(0,0,0)$ to  $(0.5,0.5,0)$), the paramagnon energy shows a marked softening with doping, with a shift of up to $\Delta E\sim150$ meV at the midpoint of the zone for the highest doping, which is about half of its initial value in the antiferromagnetic phase.  
Calculations of the dynamical spin structure factor for a one-band Hubbard model, including hopping terms up to the third-nearest neighbor ($t^{\prime\prime}$), explain this behavior and accurately reproduce the observed wave vector dependence and directional variation in the Brillouin zone.
\end{abstract}

\maketitle

\section{Introduction}

Since the discovery of high-temperature superconductivity in cuprates~\cite{Bednorz1986}, the mechanism leading to Cooper pairing remains one of the most challenging mysteries in condensed matter physics. 
Early on, spin fluctuations were suggested as a possible glue~\cite{Scalapino1995329,Moriya_2003}, but no consensus has been reached so far. 
On the one hand, it has been shown that $d$-wave pairing could be mediated by spin fluctuations~\cite{Kirtley-NatPhys2006}.
Models based on the Eliashberg formalism~\cite{eliashberg1960} adapted to spin fluctuations, and parameterized from experimental data, predict large critical temperatures $T_c$ compatible with the known values for cuprates~\cite{LeTacon_nphys2011,dahm2009}. 
On the other hand, theories generally rely on uncontrolled approximations which are questionable in the context of strongly correlated systems like the cuprates. 
For instance, while Eliashberg-like models of spin fluctuations for the Hubbard model may reproduce well the weak-coupling pseudogap, important deviations are observed in the strong-coupling regime~\cite{simkovich2024}. 
A similar approach has shown that spin fluctuations alone cannot account for the pairing in the Hubbard model~\cite{dong2022}.

A pivotal issue is the doping dependence of the spin fluctuations in cuprates~\cite{LeTacon_nphys2011,LeTaconPRB2013, Dean_nmaterials2013, dean-bisco-PRL, Guarise2014, Dean20153, PengPRB2018, Lebert_2023}.
Given the quasi-two-dimensional nature of the cuprate, the focus is set on the CuO$_2$ plane and wave vectors $(h,k)$ within the corresponding two-dimensional Brillouin zone.
Resonant inelastic x-ray scattering (RIXS) easily accesses the energy scale of the paramagnon excitations at the zone boundary $X$ $(0.5,0)$, even in samples with volumes too small for inelastic neutron scattering \cite{Dean20153}. Using RIXS, it was then possible to show that the paramagnon band-width remains constant upon hole doping~\cite{LeTacon_nphys2011,LeTaconPRB2013,Dean_nmaterials2013, jia2014a, Meyers2017, Lebert_2023}. 
In contrast, a clear softening is observed at the middle of the magnetic Brillouin zone $(0.25,0.25)$ \cite{Guarise2014, Meyers2017, Ivashko2017, PengPRB2018, RobartsPRB2019}. 
Polarization analysis confirmed that the features attributed to paramagnons in the RIXS spectra were indeed related to spin-flip excitations~\cite{Minola_PRL2015, Dellea2017, PengPRB2018, Li2024}.
Yet, such thorough analysis has only been carried out on a few hole-doped cuprates.
It is important to study the universality of these results across different cuprate families. 

From the theoretical perspective, solutions of the two-dimensional single- and three-band Hubbard models using determinant quantum Monte Carlo (DQMC) have been able to capture the absence of softening of the paramagnon energy at $X$~\cite{jia2014a,Dellea2017,PengPRB2018}. 
Yet, they predict a somewhat smaller softening compared to experiments at the zone center $(0.25,0.25)$~\cite{jia2014a}.
In a previous work dedicated to \NaCCOC{} (Na-CCOC)~\cite{Lebert_2023}, some of the authors have shown that the paramagnon spectrum in the undoped compound can be reproduced accurately with a solution of the Hubbard model obtained from antiferromagnetic cluster dynamical mean-field theory (CDMFT) on small $2\times2$ clusters. 
However, already in the underdoped regime at $x=0.10$, the antiferromagnetic (AFM) order melts and CDMFT predicts a softening of the paramagnon energy at $X$, in contradiction to experiments~\cite{Lebert_2023}.
Nevertheless, CDMFT performed with an 8-site cluster allows us to reproduce quantitatively the waterfall and kink spectral anomalies in cuprates both in the undoped and hole-doped compounds, which are a consequence of the electron-(para)magnon coupling~\cite{bacq-labreuilUniversalWaterfallFeature2025}.
This raises the question of the length scale of non-local magnetic correlations needed to account for the RIXS measurements.

\begin{figure}
    \centering
    \includegraphics[width=0.9\linewidth]{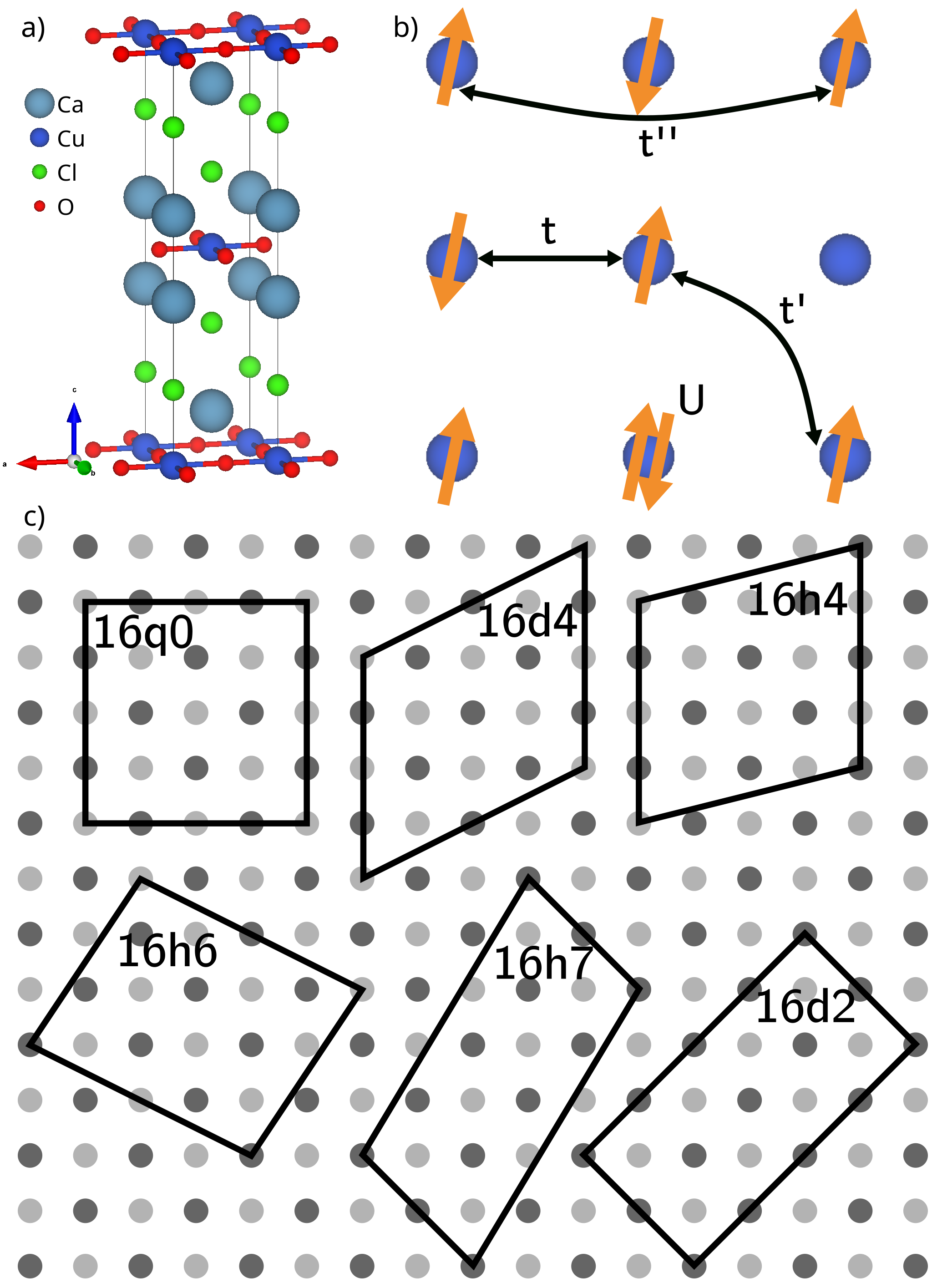}
    \caption{(a) Tetragonal crystal structure of \CCOC. (b) Effective one-band Hubbard model for the CuO$_2$ layers. (c)~The six 16-site Betts clusters used in the dynamical spin structure factor calculations (see~\cite{bettsImprovedFinitelatticeEstimates1999, bettsImprovedFinitelatticeMethod1996}). Dark and light grey represent the two sublattices of the Cu square lattice in the a-b plane, corresponding to magnetic sublattices for undoped CCOC.}
    \label{fig:structure}
\end{figure}

In this work, we study the magnetic excitations of under- ($x=0.10$) and maximally doped ($x=0.18$) Na-CCOC using RIXS. 
Na-CCOC has a simple tetragonal structure $I4/mmm$~\cite{Baptiste:wm4087} and no known structural phase transitions with doping or temperature (Fig.~\ref{fig:structure}(a)).
Its ideal electronic structure features a well-separated CuO$_2$ single band at the Fermi level~\cite{bacq2025_prx}, facilitating comparison between experiment and theories based on the single-band Hubbard model. 
We perform an exhaustive three-step analysis of the RIXS spectrum to unambiguously identify the magnetic signal.
First, a cross-polarization configuration on underdoped Na-CCOC confirms that only the spin-flip excitations contribute to the RIXS magnetic signal, and confirms the absence of softening at the zone boundary $X$. 
This complete polarization analysis also provides a justification for the use of a less time-consuming protocol exploiting a grazing geometry, which helps to disentangle the spin-flip from the non-spin-flip channels \cite{Li2024}.
The latter geometry is then leveraged to track the spin-flip excitations at specific points in the Brillouin zone both in the $(h,0)$ and $(h,h)$ directions, without further polarization analysis but with a fine $h$-mesh, and for both doping values. 
We confirm the absence of softening at $X$ even for the maximally doped sample, in contrast to an enhanced softening at the zone center $(0.25,0.25)$. 
Having justified the identification of magnetic excitations in the RIXS signal, we finally perform standard RIXS measurements to obtain a detailed paramagnon dispersion, with even finer $h$-mesh along the $(h,0)$ and $(h,h)$ directions.
Calculations of the dynamical spin structure factor, which can be used to approximate the RIXS cross section in a cross-polarized geometry \cite{jia2014a} and which are computed here via exact diagonalization on finite-size clusters of a one-band Hubbard model, reproduce the experiments accurately.  
They capture the strong softening along the $(h,h)$ direction and the absence of softening along $(h,0)$, as well as the strong broadening of the magnetic excitations.
Additionally, we observe a non-spin-flip residual at the maximum doping, at the same position as the bimagnon measured by RIXS in the undoped material~\cite{Lebert_2023}. This is in contrast to Raman measurements, which instead show a clear softening of the bimagnon energy~\cite{ChaficRamanCCOC}.

\section{Methods}
\subsection{Sample growth and preparation}
We study Na-CCOC at two doping levels: an underdoped ($x=0.10$, labeled UD2) sample with a superconducting critical temperature $T_c = 13 \pm 1$ K, and a maximally doped ($x=0.18$, labeled MD) sample with $T_c = 27 \pm 1$ K.
The underdoped sample is the same one used in the previous measurements published by Lebert \textit{et al.} in Ref.~\onlinecite{Lebert_2023}, where the sample synthesis is also described.

To obtain the $x=0.18$ sample, we synthesized a single crystal of Na-doped \CCOC~ using the flux method under high-pressure/high-temperature conditions. 
A Pt capsule was filled with a mixture of \CCOC, NaClO$_4$ (flux, oxidizer, and Na source), and NaCl (flux and Na source), and then put into the (Mg, Ca)O pressure-transmitting medium with a graphite furnace.  The medium was compressed to 8 GPa using a Walker-type high-pressure apparatus, available at the Department of Materials Science of the Graduate School of Engineering at Osaka Prefecture University. The sample was heated to 1523 K in 30 min, held at this temperature for 30 min, cooled to 1073 K over 480 min (8 hours), and quenched to room temperature. After cooling, the pressure was released.

To determine the doping of the superconducting samples, the samples' superconducting critical temperature onset T$_c$ was measured using a commercial VSM-SQUID MPMS3 magnetometer by Quantum Design\copyright.

The crystalline quality of the samples was checked, and the orientation of the facets was determined, using a 4-circle diffractometer Bruker D8 Venture using a Mo anode source and equipped with k-geometry goniometers, as well as with a Photon III detector. 

\subsection{RIXS measurements and data analysis}
The RIXS measurements were performed at the ERIXS endstation of the ID32 beamline \cite{Brookes2018} at the ESRF synchrotron. The incident energy was tuned to be equal to the copper L$_3$ edge ($E \sim 930$ eV). The combined energy resolution ranged between 45 and 60 meV. The sample facets were visually pre-aligned on the sample holder at the home institute, and more precise alignment was performed relative to the ERIXS spectrometer axis using the 4-circle diffractometer in its sample chamber.

For $x=0.10$, the measurements were carried out at a temperature of $T = 25 \, K$ (about 12 K above the $T_c = 13 \pm 1\, K)$, while for $x=0.18$ they were carried out at $T=35 K$ (around 8 K above the $T_c = 27 \pm 1 K)$.

To study the paramagnon, we used three instrument configurations. The first two separate the \SF\ from the \NSF\ channels, either using a polarimeter or exploiting a dependence of the cross section on the incidence angle.
Although time-consuming, they allow us to unambiguously demonstrate that specific spectral features are indeed related to the paramagnon. 
This enables us to use a third, less time-consuming standard configuration to measure the paramagnon dispersion with fine-grained reciprocal-space sampling, but without discrimination between the \SF\ and \NSF\ channels. 
In all cases, we work in the approximation that the out-of-plane dispersion is negligible, which is common for these experiments in cuprates \cite{Dean20153}. 
We only consider the projection of the full wave vector onto the $ab$ plane, noted as $q_{\parallel}$. 
It is calculated based on the spectrometer rotation with respect to the crystallographic axis. 
This is done while the scattering angle $2\theta$ is kept either at the maximum value, in order to maximize the modulus of the wave vector \cite{Dean20153}, or adapted to the rotation of the sample, in order to have a grazing-out configuration as discussed in Ref.~\onlinecite{Li2024}. 

\begin{figure*}[!htb]
   \centering
   \includegraphics[width=0.9\textwidth]{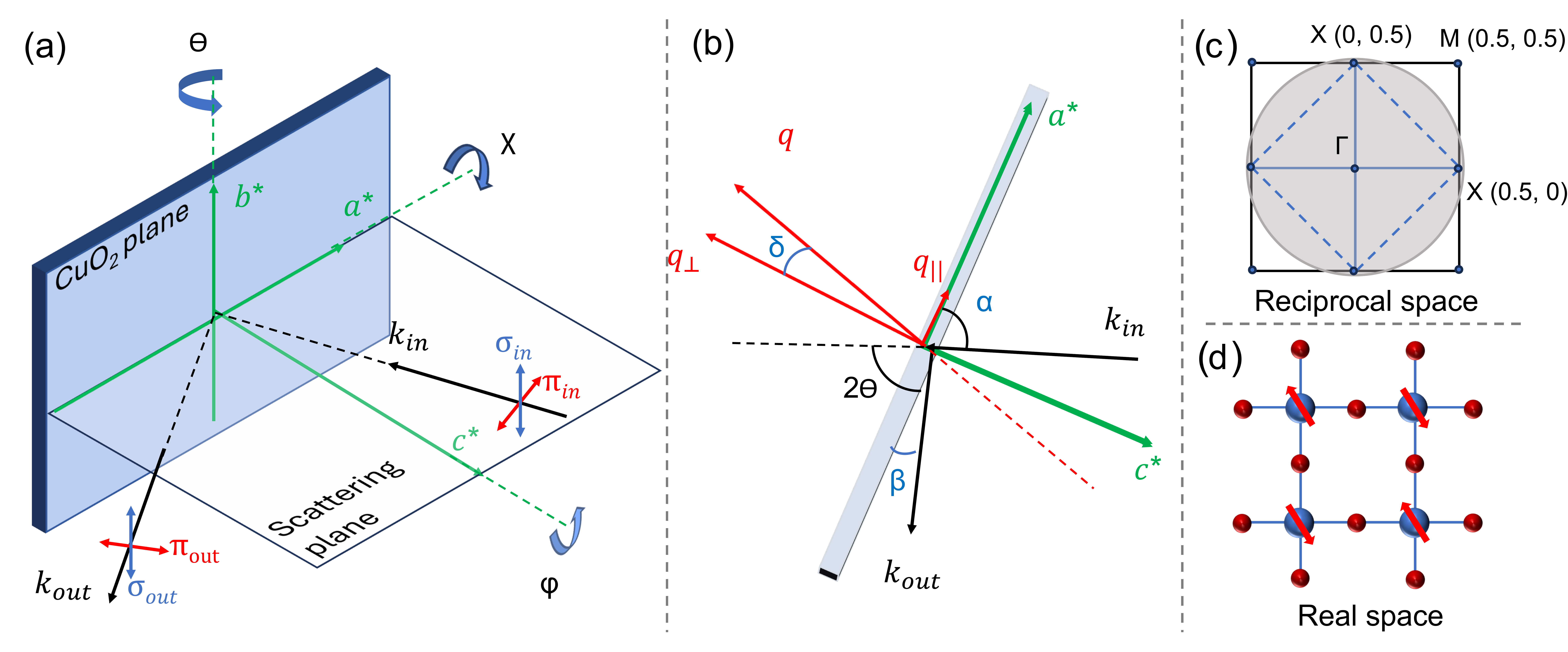}
   \caption{(a) Side view of RIXS geometry, presenting the scattering plane with respect to the CuO$_2$ planes and the incident and outgoing polarization. (b) Top view of RIXS geometry presenting the incident and outgoing wave vector ($k_{in}$ and $k_{out}$), the transferred momentum $q$ and its components ($q_{\parallel}$, and $q_{\perp}$), as well as the diffraction angle ($2\theta$), and the incident(outgoing) angle $\alpha$ ($\beta$). (c) shows reciprocal space, where the solid line represents the crystalline Brillouin zone, the dashed line represents the magnetic Brillouin zone, and the grey area represents all accessible wave-vector values in \CCOC\ at the Cu L$_3$ edge. (d) represents the square lattice of the CuO$_2$ planes.}
   \label{experiments_geometry}
\end{figure*}

\subsubsection{Polarimeter}

The first configuration used in this work exploits the $\sigma$ and $\pi$ polarizations of both the incident and scattered light to completely separate the \SF\ and \NSF\ signals \cite{braicovich2014, Jia2016, Fumagalli2019, betto2021, Li2024}. 
The selection rules for the spin configuration of \CCOC\ are presented in Table~\ref{main:pol_table2}.
For instance, the combination $\pi_{in}-\sigma_{out}$ allows to measure the \SF\ channel only (i.e., the paramagnon), while $\pi_{in}-\pi_{out}$ yields the \NSF\ channel. 
This allows the use of "simple" fitting functions consisting of one component for magnetic excitations: either a Damped Harmonic oscillator for the \SF\ signal or a Gaussian for the \NSF\ signal.
In this configuration, the spectrometer arm was fixed at its maximum value $2\theta = 149.5^{\circ}$, and we rotated the sample by changing the angle $\delta$ to change the in-plane component $q_{||}$ (see Fig.~\ref{experiments_geometry})

The price to pay for the complete separation of the \SF\ and \NSF\ channels with a polarimeter is a weak flux of scattered photons with specific polarization, thus requiring a significant acquisition time. 
Typically, the use of the polarimeter results in a reduction of the intensity down to $\sim10\%$ of the signal~\cite{braicovich2014}.

\begin{table}[!ht]
    \centering
    \begin{tabular}{c|c|cc}
     \hline \hline
                                    &\diagbox{Out}{In}& $\sigma_{in}$ & $\pi_{in}$ \\
     \hline \hline
      Spin-flip                     & $\sigma_{out}$    & 0 & $\sin^2\alpha $ \\
      $I_{\Delta S = 1}(\theta_s = 90^\circ)$  & $\pi_{out}$       & $\sin^2 \beta$  & 0\\
      \hline
      Non-spin-flip                 & $\sigma_{out}$   & 4 & 0\\
      $I_{\Delta S = 0}(\theta_s = 90^\circ)$  & $\pi_{out}$      & 0 & 4 $\sin^2 \alpha \sin^2 \beta$\\
      \hline \hline
    \end{tabular}
    \caption{Selection rules for spin-flip and non-spin-flip cross-section, depending on incident and outgoing polarization, for the spin configuration of \CCOC. The (para)magnon contribution is in the  \SF\ channel, while the \NSF\ will include the multimagnon and charge excitations.  Details of the calculation are given in Appendix~\ref{app:Polarization_dependency}.}
    \label{main:pol_table2}
\end{table}


\subsubsection{Grazing-out}\label{grazing_method}

This configuration is an alternative to polarimetry. 
It consists of fixing the scattering angle at $\beta \sim 5 - 10^\circ$ to work in a grazing-out geometry. 
We rotated the spectrometer arm, thus $2\theta$, to keep $\beta <10^{\circ}$ (see Fig.~\ref{experiments_geometry}). 
According to the selection rules in Table.~\ref{main:pol_table2}, the \SF\ channel is dominant when using $\pi_{in}$ incident light, while the \NSF\ channel is dominant with $\sigma_{in}$. 
This is summarized in Table~\ref{main:grazing_out_character} and explained in Appendix~\ref{grazing_method}. 
Hence, the grazing-out configuration allows a partial separation of the \SF\ and \NSF\ signals without analyzing the outgoing polarization, making it more time efficient. 

The separation being partial, a more complex data analysis is required. 
We separate the \SF\ from the \NSF\ by measuring both the incident polarization and using a cross-fitting, as done in Ref.~\onlinecite{Li2024} and discussed in Appendix~\ref{Cross_fitting}. 
The cross-fitting is justified by the fact that, within $\sigma$ incident polarization, the dominant non-spin-flip character is much stronger than the \SF\ one, so it can be considered a pure \NSF\ signal for practical purposes.

\begin{table}[!htb]
    \centering
    \begin{tabular}{c|c|c}
        \hline \hline
        Incident polarization & $\sigma_{in}$ & $\pi_{in}$ \\ \hline \hline
        Dominant process & Non-spin-flip & spin-flip\\ \hline
        Spin excitation & $\Delta S = 0$  & $\Delta S = 1$ \\  \hline  \hline
    \end{tabular} 
    \caption{Dominant process in \textbf{grazing-out} configuration.}
    \label{main:grazing_out_character}
\end{table}

\subsubsection{High yield measurements}

Without separation of the \SF\ from the \NSF\ signal, while fixing the scattering angle $2\theta$ to its maximum value at $149.5^\circ$, we obtain higher yields, which allows to obtain the same statistics in a reduced time, and in turn probe more spectra in the allocated beam-time frame. This allows, for example, measuring more $q$ points in the reciprocal Brillouin Zone and achieving a finer $q$-mesh, while also probing different physical conditions, such as doping, temperature, and so on. 
To enhance the paramagnon signal, we selected the $\pi_{in}$ polarization of the incident light~\cite{Braicovich_PRB2010}. Note that, for the points that are not in grazing scattering, both $\pi_{in}$ and $\sigma_{in}$ polarizations contain \SF\ as well as \NSF, so cross-fitting is not possible in that case. And we do not analyze the scattered polarization. 
The data is studied using a "simple" fit method consisting of one component for magnetic excitations: a Damped Harmonic oscillator for the \SF\ + \NSF\ signal. 

In order to change the in-plane component $q_{||}$ of the wave vector in the first Brillouin Zone, we rotate the sample by changing the angle $\delta$  (see Fig.~\ref{experiments_geometry}).\\

The three configurations are summarized in Table~\ref{method_summary} to guide the reader through the paper. 
In the following, the names of these configurations will be \textbf{highlighted in bold}.

All the data were corrected for self-absorption, as described in Appendix~\ref{Self_absorption_correction}. Further details of the fitting procedure and the Damped Harmonic oscillator model are provided in Appendix~\ref{fit-description}.

\begin{table*}[ht!]
\begin{tabular}{c|c|c|c|c|c|c}
\hline
Method  & Advantage  & Inconvenience & Polarizations & Configuration & Analysis & Typical acquisition time   \\ \hline \hline
\multirow{3}{*}{\textbf{Polarimetry}} 
& Full separation 
& \multirow{3}{*}{Time consuming} 
& $\sigma_{in}$ , $\pi_{in}$  
& $2\theta = 149.5^{\circ}$ 
& \multirow{3}{*}{simple}    
&\multirow{3}{*}{24 hours}\\
  
& of \SF\ and
&  
& $\sigma_{out}$ , $\pi_{out}$   
& $\delta$ variable    
& \\

&  \NSF 
&  
&    
&     
&     \\ \hline
\multirow{3}{*}{\textbf{Grazing-out}} 
& \multirow{3}{*}{Time efficient} 
& Partial separation 
& $\sigma_{in}$ , $\pi_{in}$ 
& $2\theta$ variable  
& \multirow{3}{*}{cross-fitting}  
& \multirow{3}{*}{2 hours}\\

&  
& of \SF\ 
& $[\sigma+ \pi]_{out}$ 
& $\beta \sim$ 5--10$^{\circ}$ 
&  \\ 

&
& and \NSF\ 
& 
& 
&   \\ \hline
\multirow{3}{*}{\textbf{High yield}} 
& Fast, suitable 
& No separation 
& $\pi_{in}$ 
& $2\theta = 149.5^{\circ}$  
& \multirow{3}{*}{simple}  
& \multirow{3}{*}{1 hour}\\

& for dispersion 
& of \SF\  
& $[\sigma+ \pi]_{out}$ 
& $\delta$ variable 
&  \\ 

& with fin $q$-mesh
&  and \NSF\ 
&  
&
& 
&   \\ \hline
\end{tabular}
\caption{Summary of the measurement methods used in this work. $[\sigma+ \pi]_{out}$ means a mixed polarization in the scattered light. See text for details.}
\label{method_summary}
\end{table*}

\subsection{Dynamical spin structure factor calculations}

To support the RIXS measurements, we perform calculations of the dynamical spin structure factor (DSSF) $S(\mathbf{q}, \omega)$ of a one-band Hubbard model. 
The latter, derived from and parametrized by \textit{ab initio} calculations, has been shown to accurately reproduce the low-energy physics of Na-CCOC~\cite{Lebert_2023, bacq-labreuilUniversalWaterfallFeature2025}.
It also provides an accurate description of the spin fluctuations~\cite{jia2014a,stepanovQuantumSpinFluctuations2018}. 
Hence, it appears as a suitable model for studying (para-)magnon properties of Na-CCOC.

The Hamiltonian of the one-band model we consider is expressed as follows: 

\begin{widetext}
\begin{equation}
    \hat{H} = - t \sum_{\langle i,j\rangle, \sigma} \cre_{i\sigma}\anh_{j\sigma} - t' \sum_{\langle\langle i,j \rangle\rangle, \sigma} \cre_{i\sigma} \anh_{j\sigma} - t'' \sum_{\langle\langle\langle i,j \rangle\rangle\rangle, \sigma} \cre_{i\sigma} \anh_{j\sigma}  - \mu \sum_i \hat{n}_i + U \sum_i \hat{n}_{i, \uparrow} \hat{n}_{i \downarrow},\label{eq:hamiltonian}
\end{equation}
\end{widetext}

where $\langle\cdot\rangle$, $\langle\langle\cdot\rangle\rangle$, $\langle\langle\langle\cdot\rangle\rangle\rangle$ are, respectively, the nearest-neighbor (NN), next-NN and next-next-NN associated with the hopping amplitudes $t = 0.436$ eV, $t'=-0.07$ eV and $t''=0.05$ eV. 
$\mu$ is the chemical potential and $U=4.34$ eV is the Hubbard onsite interaction.

The hopping amplitudes were parametrized from a density functional theory calculation performed on CCOC~using QuantumEspresso~\cite{giannozziQUANTUMESPRESSOModular2009} with the PBE functional~\cite{perdewGeneralizedGradientApproximation1996} and a plane waves + optimized norm-conserving pseudopotentials~\cite{hamannOptimizedNormconservingVanderbilt2013a} basis. 
We fitted the majority $d_{x^2-y^2}$ band with maximally localized Wannier functions~\cite{souzaMaximallyLocalizedWannier2001, marzariMaximallyLocalizedGeneralized1997}
and kept only the first three hopping terms that are comparable in phase and amplitude with a previous parametrization of this model~\cite{Lebert_2023}. 
The value of $U$ is taken from Ref.~\cite{Lebert_2023}, where it was fitted against the measured magnon spectrum of CCOC based on this model.

The zero-temperature longitudinal DSSF is computed via exact diagonalization on finite-size clusters with the XDiag package~\cite{wietekXDiagExactDiagonalization2025}.
We restrict ourselves to computations where $n_\uparrow = n_\downarrow$, preserving the SU(2) symmetry of the Hamiltonian, which allows us to restrict the computation to the $S^z$ part of the DSSF.
It is evaluated as: 
\begin{equation}
S(\mathbf{q}, \omega) =\frac{1}{N_\kappa} \sum_\mathbf{\kappa} \sum_m \left| \left\langle \Psi_m^\mathbf{\kappa} \left| S^z_\mathbf{q} \right| \Psi_0^\mathbf{\kappa}  \right\rangle \right|^2 \delta(\omega - E_m^\mathbf{\kappa} + E_0^\mathbf{\kappa})
\end{equation}
where $S^z_\mathbf{q}$ is the Fourier transformed $z$ component of the spin operator at a momentum transfer $\mathbf{q}$, $E_m^\mathbf{\kappa}$ is the energy of a state $\Psi_m^\mathbf{\kappa}$ with a twist-angle $\mathbf{\kappa}$.

Six distinct 16-site Betts clusters~\cite{bettsImprovedFinitelatticeEstimates1999, bettsImprovedFinitelatticeMethod1996} are employed, namely the 16q0, 16d4, 16h4, 16h6, 16h7 and 16d2 clusters (see Fig.~\ref{fig:structure}), which provide complementary momentum coverage and allow us to verify the transferability of spectral features.
We use twisted boundary conditions~\cite{poilblancTwistedBoundaryConditions1991, tohyamaAsymmetryElectronicStates2004}, where we apply a local gauge transformation $\hat{c}_{i, \sigma} \to e^{i \mathbf{\kappa} \cdot \mathbf{r}_i} \hat{c}_{i, \sigma}$ to the creation/annihilation operators, allowing us to shift the discrete, allowed momenta with the twist angle $\mathbf{\kappa}$.
This allows to partially alleviate the finite-size effects by averaging over a large number of twist angles. 
Furthermore, at momentum transfer $\mathbf{q}$ shared by multiple clusters, we analyze the consistency of the DSSF to distinguish collective paramagnon modes from cluster-specific artifacts. 

The use of a finite-size cluster imposes limitations on the accessible values of doping. 
In order to stay in the $S^{z}=0$ ground state sector, electrons have to be removed in pairs, such that the smallest doping value is $12.5\%$. 
We will show calculations for the undoped and the $12.5\%$ hole-doped systems. 

Finally, to ease the comparison with experimental data, the discrete exact diagonalization poles are convoluted with a Lorentzian broadening of 0.1eV. 
More details of the computational procedure are provided in Appendix~\ref{Comput_det}.

\section{Results}

\subsection{Polarimetry analysis on underdoped \texorpdfstring{\NaCCOC}{Ca2-xNaxCuO2Cl2}, x=0.10}\label{sec:polarimetry}

\begin{figure*}[!htb]
   \centering
   \includegraphics[width=1\textwidth]{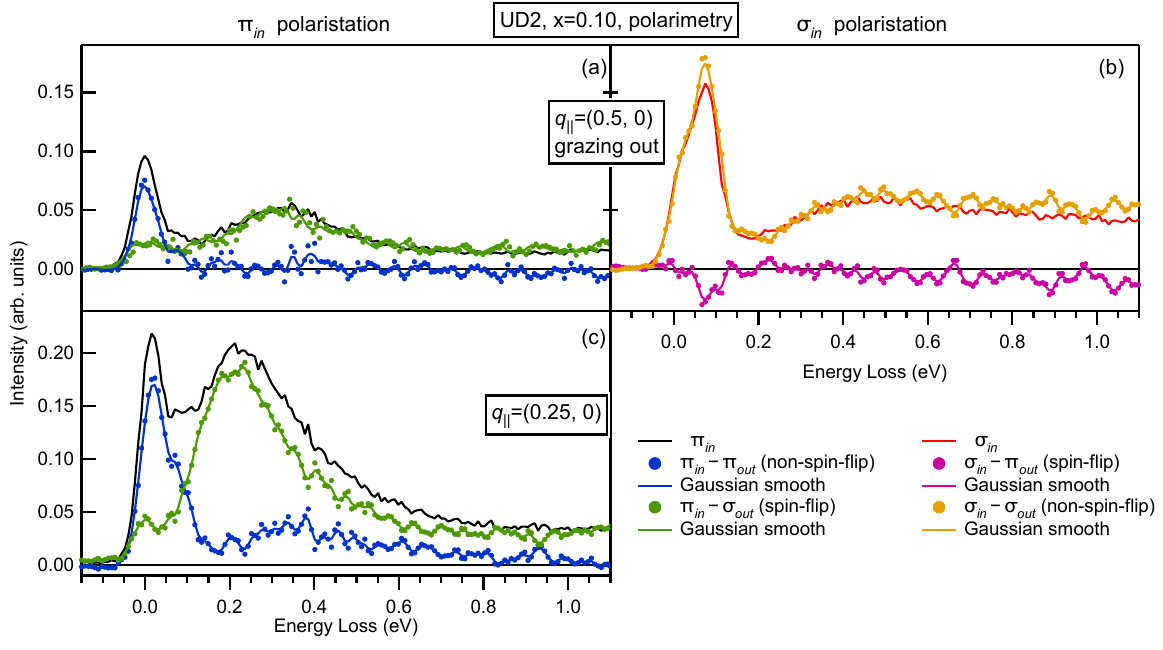}
   \caption{RIXS measurements using polarimeter along $(h,0)$ direction, at q=(0.5, 0) taken with $\pi_{in}$ (a) and $\sigma_{in}$ (b), and at q=(0.25, 0) with $\pi_{in}$ (c). The continuous black and red lines show the raw mixed-polarization data. The colored dots are the raw data of polarization-resolved spectra, while the continuous colored lines (red, green, purple, and orange) represent a Gaussian smooth.}
   \label{polarimetry_results}
\end{figure*}

The polarimeter was used to measure only two wave vectors, due to time limitations, namely: $q_{||}=(0.5, 0)$ and $q_{||}=(0.25, 0)$ for the UD2, $x=0.10$ sample. 

\paragraph{$q_{||}=(0.5, 0)$ :}
This q-vector allows us to unambiguously determine the bandwidth of the paramagnon dispersion.
In the antiferromagnetic state, the dispersion at X is in the simplest case of a Heisenberg model with nearest-neighbor super-exchange coupling $J$ given by $2J$.\\
Another interest of this point is that the scattering angle $\beta$ is about $10^{\circ}$, which implies a \textbf{grazing-out} configuration, so that the \SF\ and \NSF\ intensity can be compared to the one extrapolated from the \textbf{grazing-out} configuration without polarimetry (see section~\ref{GO_measurement}), and therefore confirm the results obtained in the latter configuration.

Figure~\ref{polarimetry_results}(a) represents the measurement with $\pi_{in}$ polarization.
One should obtain the pure \SF\ channel (the paramagnon) with crossed polarizations ($\pi_{in} - \sigma_{out}$, green dots and line), while the \NSF\ channel is obtained with parallel polarizations ($\pi_{in} - \pi_{out}$, blue dots and line), according to Table~\ref{main:pol_table2}.\\
In this \textbf{grazing-out} geometry, the \NSF\ contribution is much weaker, in agreement with Table~\ref{main:grazing_out_character}, where it is proportional to $4 \cdot sin^2(\beta)$. Consequently, the \SF\ channel (the magnon) dominates the non-polarization-resolved spectrum, represented by the solid black line.

On the other hand, in Fig.~\ref{polarimetry_results}b, the \SF\ signal is very weak ($\sigma_{in} - \pi_{out}$), also in agreement with Table~\ref{main:pol_table2}, where the signal is proportional to $sin^2(\beta)$. The \NSF\ signal is strong and does not depend on any geometrical factor. Consequently, in \textbf{grazing-out} geometry, the \NSF\ channel, which can originate from the multimagnon and charge excitation, will mainly contribute to the non-polarization-resolved spectrum using a $\sigma_{in}$ polarization (represented by the red continuous line).

These results, which are in agreement with Table \ref{main:grazing_out_character} and used in Ref.~\onlinecite{Li2024}, confirm that the \textbf{grazing-out} geometry method can be successfully used to separate \SF\ and \NSF\ contributions without the use of a time-consuming polarisation analysis, with \textbf{grazing-out} measurements with $\pi_{in}$ being mostly sensitive to \SF\ contribution while $\sigma_{in}$ is sensitive to the \NSF\ one. (Further details are provided in Appendix~\ref{app:grazing_out_meth}).

\paragraph{$q_{||}=(0.25, 0)$ :}
For this q-vector, we only measured with $\pi_{in}$ polarization in a normal, non-grazing configuration.

The measurements are shown in Fig.~\ref{polarimetry_results}c. Here, we can notice the contributions of both the \SF\ (green dots and line) as well as a small \NSF\ signal (blue dots and line) to the total signal. 

\begin{figure*}[!htb]
   \centering
   \includegraphics[width=1\textwidth]{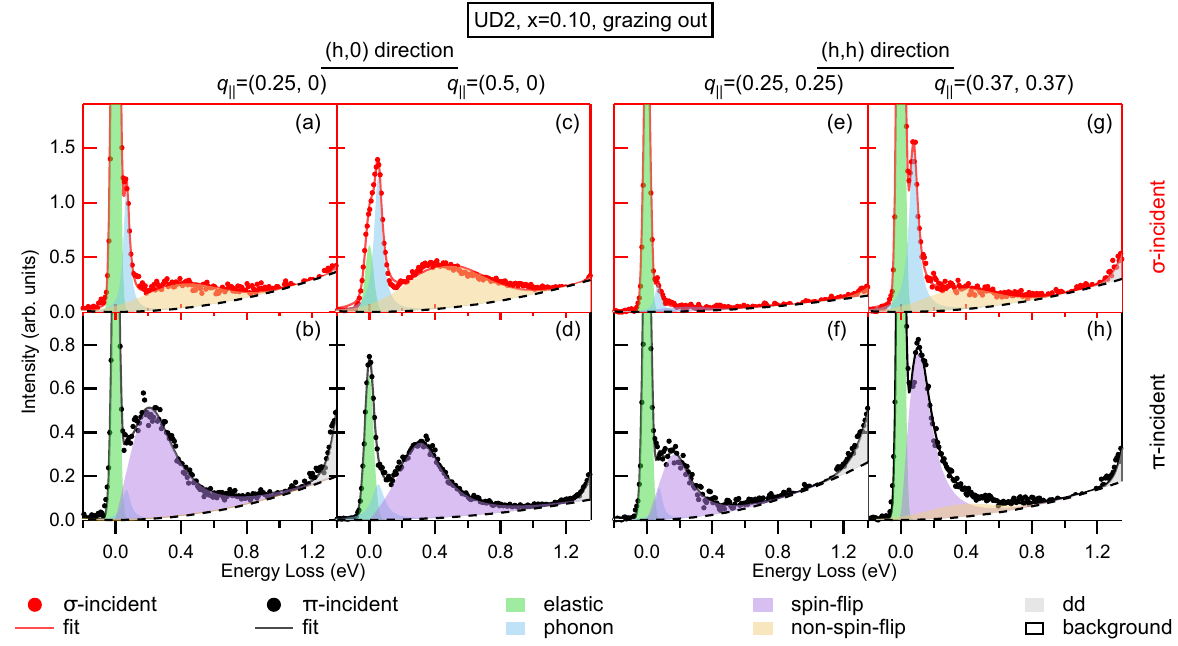}
   \caption{\textbf{Grazing-out} geometry measurements for $x=0.10$ doping, using the cross-fitting method to separate the \SF (magnon) from \NSF (multimagnon or charge excitations). The top row corresponds to $\sigma_{in}$ where the \NSF\ is dominant, while the bottom row corresponds to the measurement with $\pi_{in}$ where the \SF\ is dominant. Panels (a)--(d) show data taken along the $(h,0)$ direction, and Panels (e)--(h) show data along the $(h,h)$ direction.}
   \label{Grazing_10}
\end{figure*}

\subsection{Grazing-out measurements}\label{GO_measurement}
\subsubsection{Underdoped \texorpdfstring{\NaCCOC}{Ca2-xNaxCuO2Cl2}, x=0.10}
Having verified the validity of the \textbf{grazing-out} configuration, we can use it as an alternative to \textbf{polarimetry} measurements to separate the \SF\ (the paramagnon) from the \NSF\ (coming from the multimagnon or other charge excitations). The measurements are done with a grazing angle ($\beta$) between $5^{\circ}$ and $10^{\circ}$ on four q-vectors:
\begin{enumerate}[noitemsep, topsep=0pt]
  \item Two along the $(h,0)$ direction: $q_{||}=(0.25, 0)$ and $q_{||}=(0.5, 0)$
  \item Two along the $(h,h)$ direction: $q_{||}=(0.25, 0.25)$  and $q_{||}=(0.37, 0.37)$
\end{enumerate}
Figure.~\ref{Grazing_10} shows the measurements, with data acquired with $\sigma_{in}$ polarization in the top row and data acquired with $\pi_{in}$ polarization in the bottom row. 
The data measured with $\sigma_{in}$ were fitted without including an \SF\ contribution, as this contribution should be very weak according to Table~\ref{main:grazing_out_character}. In contrast, the $\pi_{in}$ data were analyzed using the cross-fitting method (Appendix~\ref{Cross_fitting}) to separate the \SF\ signal from the \NSF\ contribution (even though the \SF\ signal is dominant in this configuration, the \NSF\ can have a small contribution, as in Fig.~\ref{Grazing_10}(h)).

Along the $(h,0)$ direction (panels (a-b-c-d) in Fig.~\ref{Grazing_10}), there is no \NSF\ contribution in the $\pi_{in}$ channel. Along the $(h,h)$ direction (panels (e-f-g-h) in Fig.~\ref{Grazing_10}), there is a weak \NSF\ contribution only for $q_{||}=(0.37, 0.37)$ signal (Fig.~\ref{Grazing_10}(g,h)), while it is almost absent at $q_{||}=(0.25, 0.25)$ (Fig.~\ref{Grazing_10}(e,f)).

\subsubsection{Maximally doped \texorpdfstring{\NaCCOC}{Ca2-xNaxCuO2Cl2}, x=0.18}
\begin{figure*}[!htb]
   \centering
   \includegraphics[width=1\textwidth]{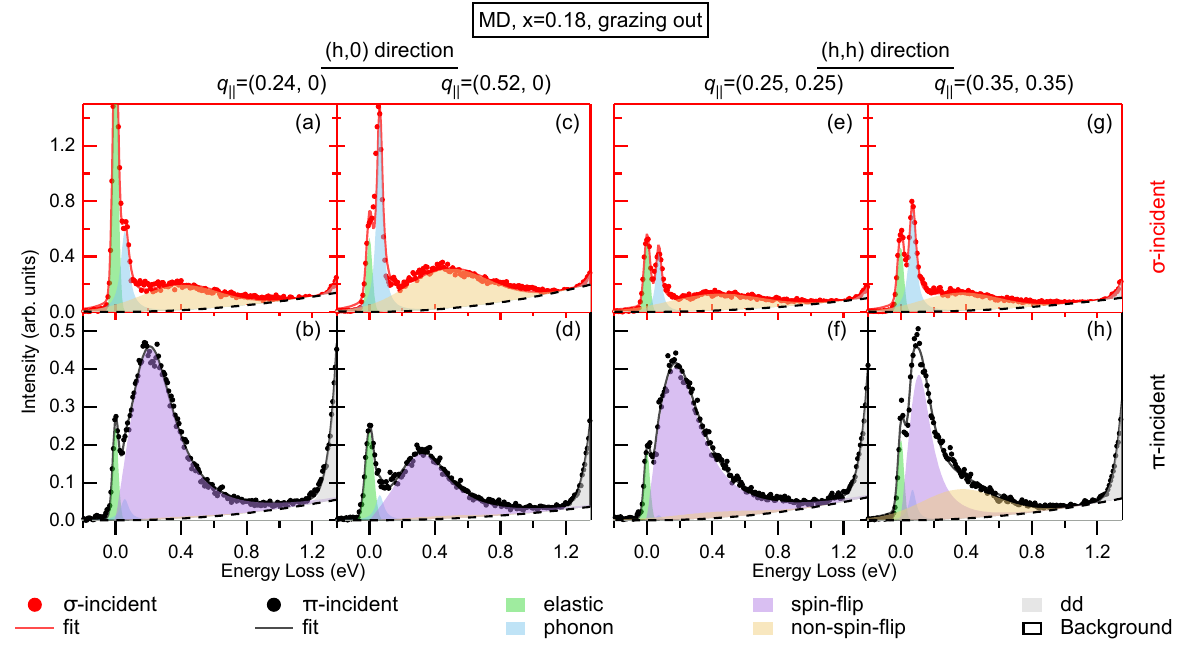}
   \caption{\textbf{Grazing-out} geometry measurements for $x=0.18$ maximum doping, with the cross-fitting method. The Top row corresponds to $\sigma_{in}$, where the \NSF\ is dominant, while the bottom row corresponds to the measurement with $\pi_{in}$, where the \SF\ is dominant. Panels (a)--(d) represent the data taken along the $(h,0)$ direction, and Panels (e)--(h) represent the data along the $(h,h)$ direction.}
   \label{Grazing_18}
\end{figure*}

The grazing-out measurements for this doping were done at the same scattering angles and the same four q-vectors as for $x=0.10$.
The measurements are presented in Fig.~\ref{Grazing_18}, where the measurements are recorded with $\sigma_{in}$ in the top row and with $\pi_{in}$ in the bottom row.
The $(h,0)$ direction shows similar results to $x=0.10$, where the \NSF\ signal does not contribute to the $\pi_{in}$ channel. In contrast, along the $(h,h)$ direction, the \NSF\ signal has a significant contribution, stronger than the one observed at $x=0.10$.

\subsection{High yield measurements with fine $q$-mesh and dispersion}
\begin{figure*}[!htb]
   \centering
   \includegraphics[width=1\textwidth]{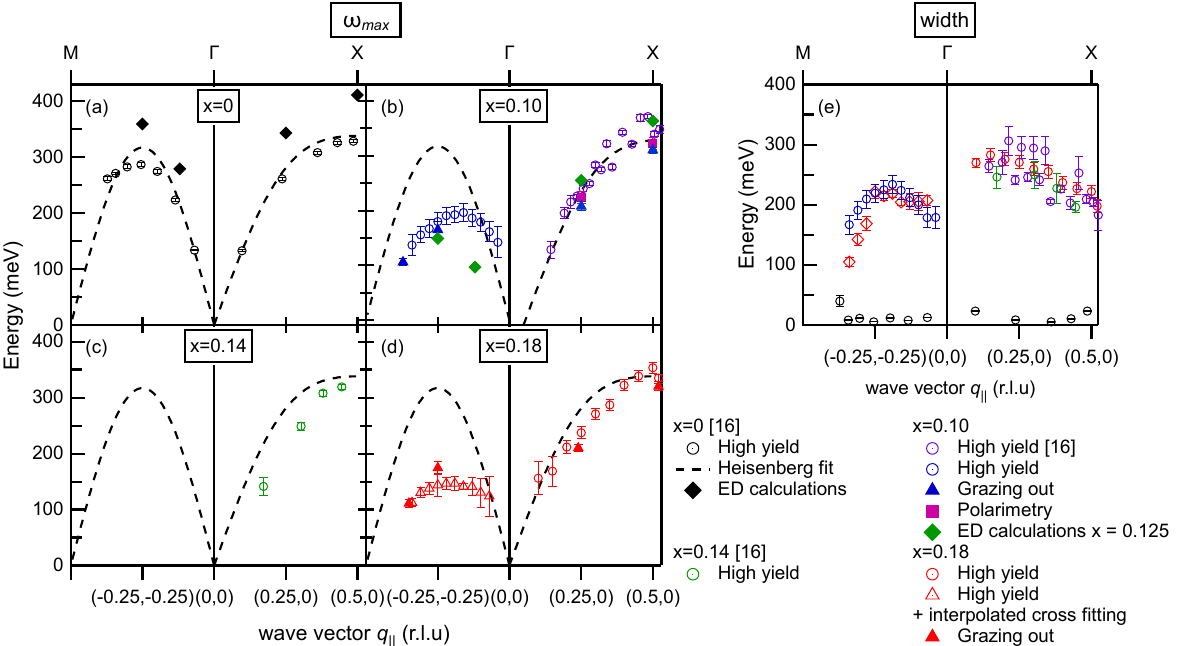}
   \caption{Doping dependence of the (para)magnon dispersion up to the maximum doping, including present and previously reported data \cite{Lebert_2023}. (a), (b), (c), and (d) correspond respectively to the $x=0$, $x=0.10$, $x=0.14$, and $x=0.18$ doping. Exact diagonalization (ED) calculations for $x=0$ and $x=0.125$ are shown in panels (a) and (b), respectively. The dashed line is the fitted Heisenberg model of the x=0 doping (AFM order) and is also plotted on the other dopings for comparison. (e) cDoping dependence of the (para)magnon width. Note that the data from \cite{Lebert_2023} were not corrected for self-absorption, unlike the data taken in this work. However, as shown in Fig.~\ref{SA_correction} in Appendix~\ref{Self_absorption_correction}, while this correction could affect the overall signal intensity, it has less impact on the peak lineshape, allowing us to reasonably compare the corresponding fitted parameters across datasets. Note also that the fit in Ref.~\onlinecite{Lebert_2023} was done using an anti-symmetrized Lorentzian, while in this work, we used a Damped Harmonic Oscillator.}
   \label{dispersion_all}
\end{figure*}

RIXS spectra measurements with \textbf{High yield} and fine $q$-mesh are recorded for $x=0.10$ along the $(h,h)$ direction, complementing the previously reported data in Ref.~\onlinecite{Lebert_2023}, and for $x=0.18$ along both the $(h,0)$ and $(h,h)$ directions. 
The complete datasets and corresponding analyses are provided in Appendix~\ref{Standard_disp}, with details of the model fitting in Appendix~\ref{fit-description}. 
From these fits, we extracted the position of the maximum of the \SF\ signal ($\omega_{max}$) and its width.
Figure~\ref{dispersion_all} shows these fitted parameters, along with those obtained for the RIXS spectra measured using \textbf{polarimetry} and \textbf{grazing-out} experimental configurations, which allow unambiguous disentanglement of the \SF\ signal. Overall, the fitted $\omega_{max}$ values for the \textbf{high yield} signal match well with those from the other configurations, validating our proposed methodology and fitting model, even when both \SF\ and \NSF\ channels contributed to the RIXS signal. 
For the \SF\ contribution, our model uses a Damped-Harmonic-Oscillator (DHO), which has been shown to better accounts for the paramagnon lineshape~\cite{Lamsal2016,PengPRB2018}, particularly in the doped regime, when the signal is damped . 
Note finally that we chose to extract the position of the maximum $\omega_{max}$, rather than the DHO propagation energy $\omega_q$, as detailed in Appendix~\ref{fit-description}, so it can be directly compared to our Dynamical Spin Structure Factor calculations, as detailed in the next section.

\subsection{Dynamical Spin Structure Factor calculations}
\begin{figure}
    \centering
    \includegraphics[width=0.95\linewidth]{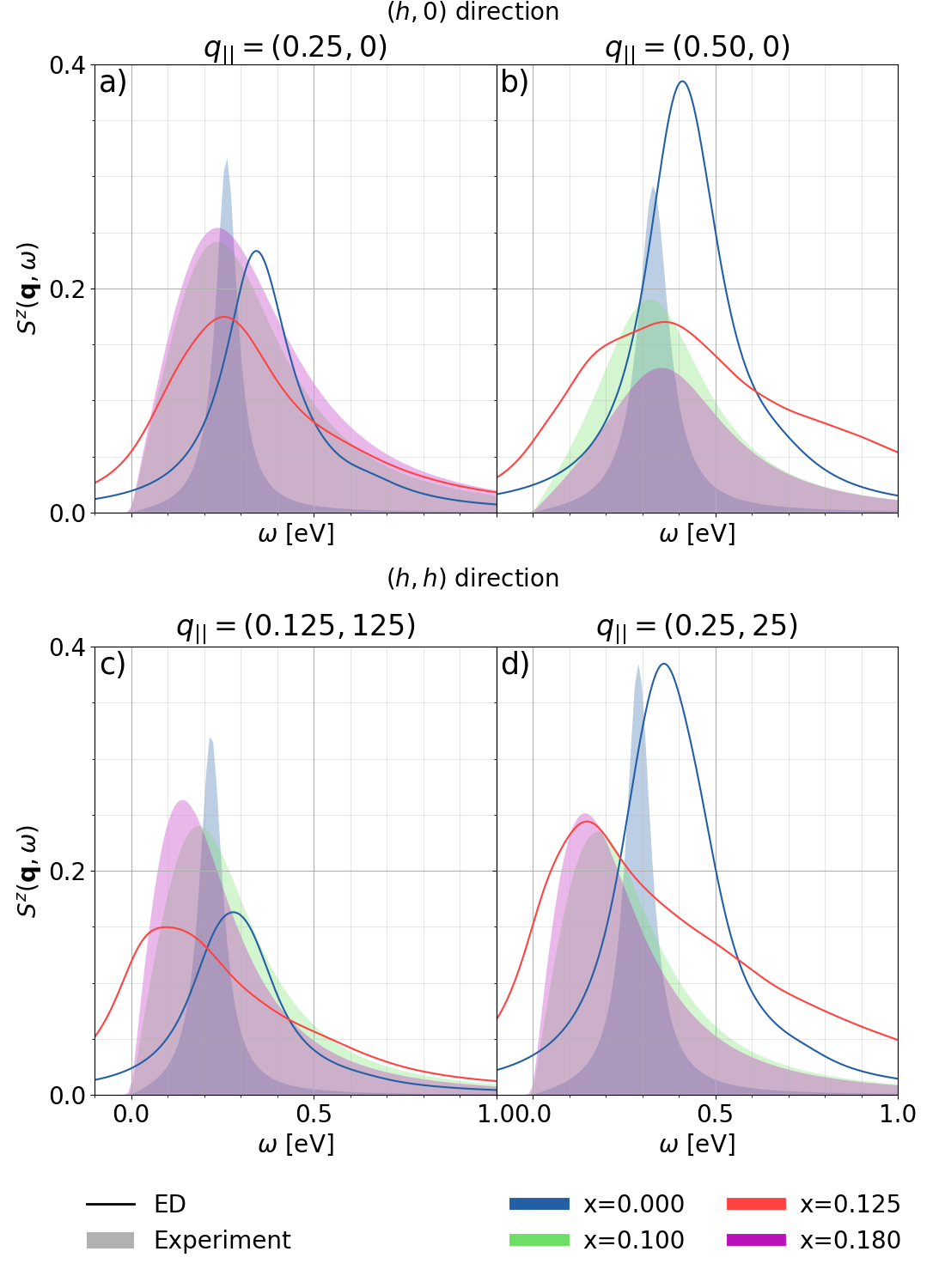}
    \caption{Comparison between the fits of the (para)magnon peak in RIXS spectra and the calculated dynamical spin structure factor. (a)-(d) DSSF at the momentum transfer points $(0.25, 0.0)$, $(0.5, 0.0)$, $(0.125, 0.125)$ and ($0.25, 0.25$) respectively. Shaded areas are fits of the (para)magnon peaks extracted from the experimental data for $x=0$, $x=0.10$, and $x=0.18$. Solid lines are the calculated spectra averaged over the different twist angles and clusters for $x=0$ and $x=0.125$.}
    \label{fig:dssf}
\end{figure}

We compare the experimental spin-flip RIXS spectra with the computed DSSF at representative momentum transfers along both high-symmetry directions. 
The theoretical spectra shown in Fig.~\ref{fig:dssf} are obtained by averaging over clusters that contain the targeted momentum point, which allows us to distinguish robust collective features from cluster-specific finite-size artifacts. 
A cluster-by-cluster comparison is provided in Appendix~\ref{Comput_det}. 

Along the anti-nodal direction $\Gamma-X$, the calculation reproduces the absence of softening at the zone boundary $X$-point $\mathbf{q} = (0.5, 0)$ (Fig.~\ref{fig:dssf}b), consistent with the polarimetry presented in Appendix~\ref{sec:polarimetry}. 

At the intermediate momentum $\mathbf{q} = (0.25, 0)$ (Fig.~\ref{fig:dssf}c), the theory captures a partial softening, in qualitative agreement with the experiment. 
These results confirm that the paramagnon energy along $(h, 0)$ is robust against doping at the level of the $t-t^{\prime}-t^{\prime\prime}$ one-band Hubbard model.

Along the nodal direction $\Gamma-M$, the calculation reproduces the strong softening of the paramagnon observed experimentally at $\mathbf{q} = (0.25, 0.25)$ (Fig.~\ref{fig:dssf}(a)).
Without any adjustment of the model parameters and by tuning the electron filling, the calculation naturally accounts for the marked contrast between the two high-symmetry directions, capturing the anisotropy of the doping-induced renormalization of spin fluctuations.

Whereas a further increase of the cluster sizes used in our simulation is hindered by the exponential growth of the corresponding many-body Hilbert space, we do not expect major changes of the DSSF for larger clusters.
Importantly, the 16-site clusters employed here already account for all types of hopping terms in the effective model Hamiltonian.
Moreover, since the magnetic correlation length of the Hubbard model decreases with the interaction strength $U$ in the strong-coupling regime and becomes rather short-range \cite{simkovic2022}, we believe that salient features of the DSSF are already captured with the current cluster sizes.

\section{Discussion}

\subsection{The paramagnon}
Figure~\ref {dispersion_all} shows the doping dependence of the paramagnon dispersion, with all data taken in the \textbf{high yield} configuration, \textbf{grazing-out} configuration, and \textbf{polarimetry} data. The paramagnon width for \textbf{high yield} measurements is plotted in Fig.~\ref{dispersion_all}(e).

The paramagnon dispersion shows two main features:\\

1- A persistent energy dispersion along the $(h,0)$ direction upon doping, compared with the undoped compound. 
In undoped CCOC, the magnetic part of the Hamiltonian can be approximated by a Heisenberg model with leading super-exchange coupling $\mathcal{J}$.
Interestingly, along the $(h,0)$ direction, dominated by nearest-neighbor exchange within this simple Heisenberg approximation, doping has little effect on the dispersion, in agreement with all other hole-doped cuprates~\cite{LeTacon_nphys2011, LeTaconPRB2013, Dean_nmaterials2013, Minola_PRL2015, Meyers2017, PengPRB2018, wang2022}. 
However, a very strong broadening is observed, indicating a much shorter lifetime, which already sets in at $x=0.10$ and stays relatively constant at higher doping, as shown in Fig.~\ref{dispersion_all}(e).
Since long-range antiferromagnetic order breaks down already at ~2\% doping, one might at first expect a softening of the spin excitations \cite{Manousakis2011}.
Also, a weakening of short-range antiferromagnetic correlations could be expected based on the dilution of AFM bonds \cite{jia2014a}.
However, the picture of localized static moments breaks down in the hole-doped system, and a model of itinerant electrons such as the Hubbard model has to be considered to understand the spin dynamics \cite{Guarise2014}.
\\

2- The paramagnon softens along the $(h,h)$ direction across doping. 
When describing the paramagnons based on correlated itinerant electrons in terms of the spin susceptibility in the random phase approximation (RPA), a notable modification of the particle-hole response function occurs only close to the AFM wave vector \cite{Guarise2014}.
This leads to an anisotropic softening along the nodal direction, whereas the anti-nodal paramagnon dispersion remains largely unchanged, in qualitative agreement with our measurements here.
Although the paramagnon dispersion along the $(h,h)$ direction was not systematically investigated in RIXS, most of the previous studies focused on the doping dependence of the dispersion along the $(h,0)$ direction. It was however reported and discussed in several studies in La$_{2-x}$Sr$_x$CuO$_4$ \cite{Meyers2017, Ivashko2017, RobartsPRB2019}, Bi$_2$Sr$_2$CaCu$_2$O$_{8+\delta}$ \cite{Guarise2014}, as well as in (Bi,Pb)$_2$(Sr,La)$_2$CuO$_{6+\delta}$ \cite{PengPRB2018}.

Along the $(h,h)$ direction, the dispersion does not peak at $q_{||} = (0.25, 0.25) $, but instead the maximum seems to shift toward a lower wave vector. This suggests a shift from the commensurate magnetic wave vector along $(h,h)$. This shift has also been observed in La$_{2-x}$Sr$_x$CuO$_4$ by Robarts \textit{et al.} \cite{RobartsPRB2019}.

The calculation of the DSSF via exact diagonalization of the one-band Hubbard model provides a means to study the paramagnon evolution upon doping. 
We should mention that, in order to reintroduce charge degrees of freedom, one could be tempted to resort to the $t$--$J$ model, which is its strong-coupling limiting case $t\ll U$~\cite{Chao1977tJ,Auerbach1994}.
In the hole-doped regime relevant to us here, however, higher order processes absent in $t-J$-type models become important for a quantitative description of the paramagnon dispersion \cite{jia2014a}, such that we focus instead on the Hubbard model, which still captures the consequences of doping in the charge channel.
Using Betts clusters with twist-averaged boundary conditions, we obtain qualitative agreement with experiment and capture the anisotropic paramagnon softening along the $(h,0)$ and $(h,h)$ directions.
As for the precise positions of the paramagnon peaks at zero doping, $x=0$, the simulations lead to slightly larger paramagnon energies at $\mathbf{q}=(0.5,0)$ and $\mathbf{q}=(0.25,0.25)$.
Upon doping, the softening with respect to the peak position at $x=0$ is then systematically overestimated in our simulations. 
In principle, this difference to the experimental data can be rooted either in the simplified description of the material's electronic structure via a one-band Hubbard model or in the subsequent approximate treatment of this model $via$ exact diagonalization.

For instance, a description of the CuO$_2$ planes explicitly including the oxygen degrees of freedom via an Emery-like model \cite{zaanen1985band,emery1987theory} could lead to a better description of the relaxation processes through the ligands. 
Combined with a precise \textit{ab initio} determination of the Coulomb interaction, in line with a description of both one- and two-body parts of the Hamiltonian at the same level of theory, it could lead to a better description of the excitation energies and finally change the amplitude of the paramagnon softening.
Indeed, given that the strong-coupling limit of this model in linear spin-wave theory already shows a slight overestimation of the magnon dispersion at zero doping, in particular at $\mathbf{q}=(0.25,0.25)$, see Fig.~\ref{dispersion_all}, part of the discrepancy at $x=0$ seems to be rooted in approximations on the model level.

Another aspect, linked to the technique of our choice to treat the Hubbard model, is finite-size effects within exact diagonalization, which show up in the form of significant charge discretization effects upon doping.
For instance, in order to stay in a $S^z=0$ ground state, the smallest realizable doping on the 16-site clusters studied here is a doping of 12.5\%, which amounts to the removal of two electrons.
Unfortunately, this doping also coincides with several instabilities towards stripe phases \cite{zheng2017stripe,huang2018stripe,xu2022stripes,xu2024coexistence}, whose inclusion might be important for a better quantitative agreement with experiment.

Finally, we would like to note that the softening of the paramagnon along the $(h,h)$ direction is altered by the inclusion of a third nearest-neighbor hopping term, $t^{\prime\prime}$, in the Hamiltonian, see Appendix~\ref{sec:3NNhop}.
This also explains the difference between our findings and previous works on the $t-t^{\prime}-U$ Hubbard model, where the paramagnon softening was much smaller \cite{jia2014a}.
In particular, the role of three-site exchange \cite{Daghofer_2008}, which was identified as an important ingredient to explain the antiferromagnetic zone boundary paramagnon dispersion \cite{jia2014a}, is here weakened:
Whereas a misaligned spin in an antiferromagnetic background reduces the mobility of a hole by removing three-site hopping terms involving the nearest-neighbor hopping $t$, both hopping processes involving $t^{\prime}$ and $t^{\prime\prime}$ do not suffer from this intermediate site blockade.
The overall effect of the three-site terms for the reduction of the hole-delocalization energy is therefore smaller than in a $t-U$ or $t-t^{\prime}-U$ Hubbard model.

\subsection{The non-spin-flip signal}

\begin{figure}[!htb]
   \centering
   \includegraphics[width=0.5\textwidth]{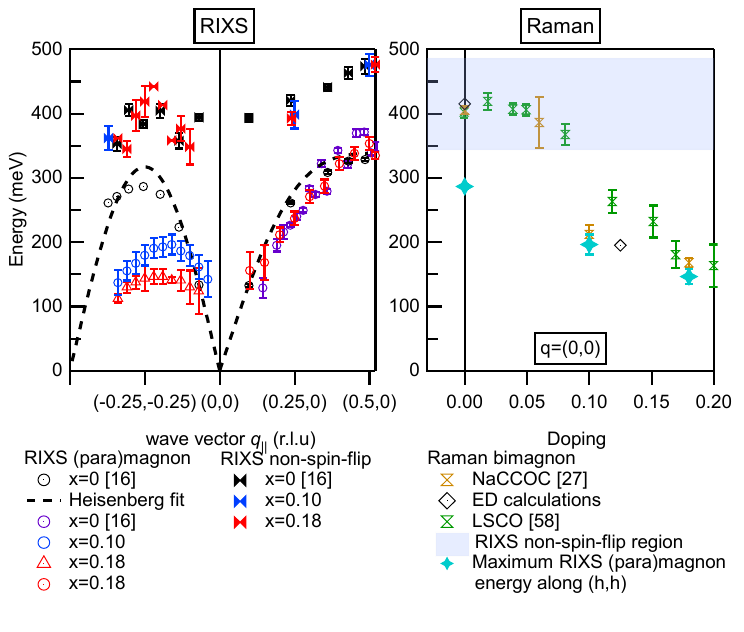}
   \caption{(Left panel)\SF\ and \NSF\ energy dispersions in the \NaCCOC\ obtained from RIXS in this work and Ref.~\onlinecite{Lebert_2023}. The paramagnons are those reported in Fig.~\ref{dispersion_all} using the \textbf{high yield} method; For $x=0.10$ and $x=0.18$, the \NSF\ energies are obtained from the \textbf{grazing-out} $\sigma_{in}$ measurements, except for $x=0.18$ along $(h,h)$, for which the energies were extracted from \textbf{high yield} $\pi_{in}$ measurements.
   (Right panel) Doping dependence of the Raman $B_{1g}$ bimagnon energy in \NaCCOC\ from Ref.~\onlinecite{ChaficRamanCCOC} at $q\sim\Gamma$, together with corresponding exact diagonalization (ED) calculations. For comparison, the Raman $B_{1g}$ bimagnon in \LSCO\ from Ref.~\onlinecite{muschler2010} is also shown. The blue shaded region indicates the range of the RIXS \NSF, defined by the minimum and maximum energy across all doping. This panel also shows the maximum paramagnon energy along $(h,h)$ for each doping, allowing direct comparison with the doping dependence of the Raman bimagnon energy. }
   \label{bimagnon_RIXS_RAMAN}
\end{figure}

In the $\sigma_{in}$ measurements, the \NSF\ signal is present in both $x=0.10$ and $x=0.18$ dopings along the $(h,0)$ direction (Fig.~\ref{Grazing_10} (a,c) and Fig.~\ref{Grazing_18}(a,c)). Along the $(h,h)$ direction, this signal is also present in both dopings, except that it is barely visible at $x=0.10$ in the $(0.25, 0.25)$ r.l.u (Fig.~\ref{Grazing_10}(e)).
In the $\pi_{in}$ measurements, the \NSF\ signal is very weak for $x=0.10$ doping (Fig.~\ref{Grazing_10}) and has little contribution, while for $x=0.18$ (the maximum doping), a clear \NSF\ contribution can be observed (Fig.~\ref{Grazing_18}).

In RIXS, a \NSF\ signal can originate not only from a multimagnon process but also from charge and lattice excitations. Multimagnon excitations include two-magnon and bimagnon processes, as well as higher-order processes with $\Delta S = 0, 2, 4...$. We refer to a "two-magnon" excitation as a two-step process in which the photon interacts with two non-interacting magnons, yielding a maximum energy of $4J$: two interacting magnons, each with a maximum of $2J$, breaking 4 AFM bonds each, 8 bonds together~\cite{jia2014a}. In contrast, we refer to a "bimagnon" as a process involving magnon-magnon interaction leading to an energy of $\sim 3 J$ \cite{devereauxILS2007, jia2014a}: where two adjacent spin-flip occurs, breaking 6 AFM bonds in total with their surrondings. Simulations of the two-magnon density of states across the Brillouin zone in Ref.~\onlinecite{forte2008, Bisogni_PRB2012} reveal continuum-like behavior, with a relatively flat branch around $4J$ and another dispersive branch, both showing substantial spectral intensity at high energies. Including magnon-magnon interactions, the maximum energy decreases \cite{forte2008, Bisogni_PRB2012, devereauxILS2007}. At the Cu $L_3$ edge, RIXS probes the lower branch of this continuum \cite{Bisogni_PRB2012}.

Raman spectroscopy is another technique sensitive to magnetic excitations \cite{Fleury_Loudon_1968, devereauxILS2007}, measured at $\Delta q \simeq 0$, and therefore probes the upper branch of the two-magnon excitations. Including magnon-magnon interaction, Raman spectroscopy probes a bimagnon excitation that appears in the $B_{1g}$ channel at $\sim 2.7J$ \cite{devereauxILS2007} in undoped cuprates.

Figure~\ref{bimagnon_RIXS_RAMAN} compares RIXS and Raman spectroscopy measurements. 
Figure~\ref{bimagnon_RIXS_RAMAN}(left panel) summarize both \SF\ (para)magnon signal, and \NSF\ dispersion from Ref.~\onlinecite{Lebert_2023} and the present work. While the \SF\ signal shows a softening only along the $(h,0)$ direction, as discussed in the previous section, the \NSF\ shows no energy variation upon doping. Figure~\ref{bimagnon_RIXS_RAMAN}(left panel) shows the Raman $B_{1g}$ signal on Na-CCOC reported by some of the present authors~\cite{ChaficRamanCCOC}, as well as the corresponding exact diagonalization (ED) simulation from the present work as described in Appendix~\ref{Raman_signal_simulation}, and the $B_{1g}$ signal on \LSCO\ from Ref.~\onlinecite{muschler2010}, which has a similar spin-exchange coupling $J$. Both $B_{1g}$ bimagnons in Na-CCOC and \LSCO\ soften upon doping, in agreement with the ED simulation.
Since the $B_{1g}$ bimagnon in the undoped sample has an energy of $\sim~2.7 J$ because it breaks 6 AFM bonds with surrounding spins, doping reduces the number of broken AFM bonds, softening the bimagnon energy. 
Moreover, the bimagnon softening shows a similar trend to the (para)magnon softening measured in RIXS along $(h,h)$, as shown in Fig.~\ref{bimagnon_RIXS_RAMAN} (right panel). This suggests that the $B_{1g}$ bimagnon arises mainly from magnon-magnon interactions involving magnons along the $(h,h)$ direction.

In the undoped sample, the \NSF\ in RIXS shows a similar energy scale to the bimagnon in Raman spectroscopy (Fig.~\ref{bimagnon_RIXS_RAMAN}), as also reported for \LSCO in Ref.~\onlinecite{chaixRIXS2018}. However, upon doping, the $B_{1g}$ bimagnon exits the energy region that corresponds to the \NSF\ signal observed in RIXS (Fig.~\ref{bimagnon_RIXS_RAMAN}(right panel)).

Therefore, while the \NSF\ in undoped cuprates is well established as a bimagnon \cite{braicovichMagneticExcitationsPhase2010, Bisogni_PRB2012, Minola_PRL2015, Minola2017, chaixRIXS2018,  singhBimagnonDispersionLa2CuO42025}, we argue that at $x=0.18$ it is no longer magnetic in nature. At $x=0.10$, it remains unclear whether the \NSF\ can still be identified as a multimagnon or is strongly suppressed; then, at higher doping, another excitation arises. 
Several experimental studies in hole-doped cuprates~\cite {Minola_PRL2015, Minola2017, Kang2019, Fumagalli2019, Zhang2022} have discussed that doping introduces additional charge carriers, potentially leading to the emergence of charge excitations. In Ref.~\cite{Minola_PRL2015, Minola2017}, the authors showed that charge excitations (resembling incoherent particle-hole excitations) emerge and contribute to the \NSF\ channel as a continuum upon doping.
It was suggested that this appears in a crossover regime beyond the maximum doping, which might relate to a reconstruction of the electronic structure at this doping.  In Ref.~\onlinecite{Kang2019}, the authors argue that in the \NSF\ channel, the charge-to-spin ratio increases with doping. In addition, theoretical studies (Ref.~\cite{Tsutsui2016}) show that the \NSF\ signal exhibits spin-like (bimagnon) behavior up to near-maximal doping, then transitions to charge-like behavior.
However, definitive conclusions about the nature of this excitation require new measurements such as detuning the energy from resonance to reproduce the behavior reported in Ref.~\cite{Minola_PRL2015}.

\section{Conclusion}

In summary, combining three experimental configurations, we demonstrate that only the spin-flip excitations contribute to the RIXS signal in the energy range usually attributed to the paramagnon. 
We show the absence of softening of the paramagnon energy at the zone boundary $X$ in Na-CCOC at all dopings. 
In contrast, we observe a strong softening of the paramagnon with doping along the nodal direction $\Gamma-M$ (\textit{i.e.}, from (0,0) to (0.5,0.5)) reaching up to $\sim 150$ meV at the middle of the zone for the maximum doping. 
Our work therefore confirms the universality of the doping dependence of the paramagnon observed in previous works~\cite{_nphys2011, LeTaconPRB2013,Dean_nmaterials2013,jia2014a,Meyers2017,Lebert_2023}.

Calculations of the dynamical spin structure factor capture all the salient experimental observations: both the softening at the zone center $(0.25,0.25)$ and the absence of softening at $X$ are well reproduced, along with the enhanced broadening with doping. 
We show that the softening of the paramagnon along the $(h,h)$ direction is significantly altered by the inclusion of a third nearest-neighbor hopping term, $t^{\prime\prime}$, in the Hamiltonian. 
As a consequence, the overall effect of the three-site terms~\cite{jia2014a} on the reduction of the hole-delocalization energy is therefore less important, as both the second and third nearest-neighbor hopping terms, $\propto t^{\prime},t^{\prime\prime}$, provide additional hole-delocalization channels.

Additionally, we observe a non-spin-flip residual, re-emerging at the maximum doping, at the same position where the bimagnon is measured by RIXS in the undoped~\cite{Lebert_2023}, antiferromagnetic compound. 
This is in contrast to Raman measurements, which instead show a clear softening of the bimagnon energy, in line with our ED simulations of the Raman $B_{1g}$ intensity that also show this softening.
The interpretation of that excitation emerging at maximum doping will require further investigation.

\section*{Acknowledgment}
RIXS measurements were done at the ESRF under proposal numbers HC-5226~\cite{esrf_experiment_5226} and HC-5427~\cite{esrf_experiment_5427}.

This project was provided with HPC computing and storage resources by GENCI at TGCC thanks to the grants 2025-GEN12043 and 2026-GEN17548 on the supercomputer Joliot Curie's ROME partition.
L.G. and B.L. acknowledge financial support from the French National Research Agency via the ANR-JCJC project ClusterSpecs (ANR-24-CE30-5387).

We thank Olivier Leynaud and the X’Press pool for help on single-crystal diffraction, as well as Oscar Ramon Fabelo Rosa for X-ray characterization of the maximally doped sample.
We thank Didier Dufeu and the PMag platform for help in the magnetometry measurements. 
We thank Pierre Toulemonde, Coraline Letouz\'e, and Nicolas Renard for help and discussions during the beamtime. 

We are very grateful to M. Le Tacon and M. P. M. Dean for an insightful and critical reading of the manuscript.

\bibliographystyle{apsrev}
\bibliography{references}

@Article{vaknin_prb,
  Title                    = {Stacking of the square-lattice antiferromagnetic planes in Ca2CuO2Cl2},
  Author                   = {D. Vaknin and L. L. Miller and J. L. Zarestky},
  Journal                  = {Phys. Rev. B},
  Year                     = {1997},
  Number                   = {13},
  Pages                    = {8351},
  Volume                   = {56},
  doi = {10.1103/PhysRevB.56.8351},
  url = {https://link.aps.org/doi/10.1103/PhysRevB.56.8351}
}

@Article{Li2024,
  title = {Evolution of the Magnetic Excitations in Electron-Doped ${\mathrm{La}}_{2\ensuremath{-}x}{\mathrm{Ce}}_{x}{\mathrm{CuO}}_{4}$},
  author = {Li, X. T. and Tu, S. J. and Chaix, L. and Fawaz, C. and d'Astuto, M. and Li, X. and Yakhou-Harris, F. and Kummer, K. and Brookes, N. B. and Garcia-Fernandez, M. and Zhou, Ke-Jin and Lin, Z. F. and Yuan, J. and Jin, K. and Dean, M. P. M. and Liu, X.},
  journal = {Phys. Rev. Lett.},
  volume = {132},
  issue = {5},
  pages = {056002},
  numpages = {7},
  year = {2024},
  month = {Feb},
  publisher = {American Physical Society},
  doi = {10.1103/PhysRevLett.132.056002},
  url = {https://link.aps.org/doi/10.1103/PhysRevLett.132.056002}
}

@Article{Bednorz1986,
  Title                    = {Possible highTc superconductivity in the Ba−La−Cu−O system},
  Author                   = {Bednorz, J. G. and M{\"u}ller, K. A.},
  Journal                  = {Zeitschrift für Physik B Condensed Matter},
  Year                     = {1986},
  Number                   = {2},
  Pages                    = {189--193},
  Volume                   = {64},
  Doi                      = {10.1007/BF01303701},
  ISSN                     = {1431-584X},
  Refid                    = {Bednorz1986},
  Url                      = {https://doi.org/10.1007/BF01303701}
}

@Article{Kirtley-NatPhys2006,
  Title                    = {Angle-resolved phase-sensitive determination of the in-plane gap symmetry in YBa2Cu3O7−δ},
  Author                   = {Kirtley, J. R. and Tsuei, C. C. and Ariando and Verwijs, C. J. M. and Harkema, S. and Hilgenkamp, H.},
  Journal                  = {Nat. Phys.},
  Year                     = {2006},
  Number                   = {3},
  Pages                    = {190--194},
  Volume                   = {2},
  ISSN                     = {1745-2481},
  Refid                    = {Kirtley2006},
  Url                      = {https://doi.org/10.1038/nphys215}
}

@Article{wang2022,
  Title                    = {Paramagnons and high-temperature superconductivity in a model family of cuprates},
  Author                   = {Wang, Lichen and He, Guanhong and Yang, Zichen and Garcia-Fernandez, Mirian and Nag, Abhishek and Zhou, Kejin and Minola, Matteo and Tacon, Matthieu Le and Keimer, Bernhard and Peng, Yingying and Li, Yuan},
  Journal                  = {Nat. Commun.},
  Year                     = {2022},
  Number                   = {1},
  Pages                    = {3163--},
  Volume                   = {13},
  ISSN                     = {2041-1723},
  Refid                    = {Wang2022},
  Url                      = {https://doi.org/10.1038/s41467-022-30918-z}
}

@article{Guarise2014,
  title = {Anisotropic Softening of Magnetic Excitations along the Nodal Direction in Superconducting Cuprates},
  author = {Guarise, M. and Piazza, B. Dalla and Berger, H. and Giannini, E. and Schmitt, T. and R{\o}nnow, H. M. and Sawatzky, G. A. and Van Den Brink, J. and Altenfeld, D. and Eremin, I. and Grioni, M.},
  journal = {Nat. Commun.},
  year = {2014},
  month = {dec},
  volume = {5},
  number = {1},
  pages = {5760},
  issn = {2041-1723},
  doi = {10.1038/ncomms6760},
  url = {https://doi.org/10.1038/ncomms6760},
}

@article{Manousakis2011,
  title = {Hole spectral functions in lightly doped quantum antiferromagnets},
  author = {Kar, Satyaki and Manousakis, Efstratios},
  journal = {Phys. Rev. B},
  volume = {84},
  issue = {20},
  pages = {205107},
  numpages = {12},
  year = {2011},
  month = {Nov},
  publisher = {American Physical Society},
  doi = {10.1103/PhysRevB.84.205107},
  url = {https://link.aps.org/doi/10.1103/PhysRevB.84.205107}
}

@article{Lebert_2023,
  title = {Paramagnon dispersion and damping in doped {${\mathrm{Na}}_{x}{\mathrm{Ca}}_{2\text{\ensuremath{-}}x}{\mathrm{CuO}}_{2}{\mathrm{Cl}}_{2}$}},
  author = {Lebert, Blair W. and Bacq-Labreuil, Benjamin and Dean, Mark P. M. and Ruotsalainen, Kari and Nicolaou, Alessandro and Huotari, Simo and Yamada, Ikuya and Yamamoto, Hajime and Azuma, Masaki and Brookes, Nicholas B. and Yakhou, Flora and Miao, Hu and Santos-Cottin, David and Lenz, Benjamin and Biermann, Silke and d'Astuto, Matteo},
  journal = {Phys. Rev. B},
  volume = {108},
  issue = {2},
  pages = {024506},
  numpages = {10},
  year = {2023},
  month = {Jul},
  publisher = {American Physical Society},
  doi = {10.1103/PhysRevB.108.024506},
  url = {https://link.aps.org/doi/10.1103/PhysRevB.108.024506},
}

@Article{bacq-labreuilUniversalWaterfallFeature2025,
  author    = {Bacq-Labreuil, Benjamin and Fawaz, Chafic and Okazaki, Yuichi and Obata, Yukiko and Cercellier, Herv\'e and Le F\`evre, Patrick and Bertran,  Fran{\ifmmode \mbox{\c{c}}\else \c{c}\fi{}}ois and Santos-Cottin, David and Yamamoto, Hajime and Yamada, Ikuya and Azuma, Masaki and Horiba, Koji and Kumigashira, Hiroshi and d'Astuto, Matteo and Biermann, Silke and Lenz, Benjamin},
  journal   = {Phys. Rev. Lett.},
  title     = {Universal Waterfall Feature in Cuprate Superconductors: Evidence of a Momentum-Driven Crossover},
  year      = {2025},
  month     = {Jan},
  pages     = {016502},
  volume    = {134},
  doi       = {10.1103/PhysRevLett.134.016502},
  issue     = {1},
  numpages  = {8},
  publisher = {American Physical Society},
  url       = {https://link.aps.org/doi/10.1103/PhysRevLett.134.016502},
}

@Article{Moriya_2003,
  Title                    = {Antiferromagnetic spin fluctuation and superconductivity},
  Author                   = {T{\^o}ru Moriya and Kazuo Ueda},
  Journal                  = {Reports on Progress in Physics},
  Year                     = {2003},

  Month                    = {jul},
  Number                   = {8},
  Pages                    = {1299--1341},
  Volume                   = {66},
  Doi                      = {10.1088/0034-4885/66/8/202},
  Publisher                = {{IOP} Publishing},
  Url                      = {https://doi.org/10.1088/0034-4885/66/8/202}
}

@Article{Scalapino1995329,
  Title                    = {The case for dx2-y2 pairing in the cuprate superconductors},
  Author                   = {D.J. Scalapino},
  Journal                  = {Physics Reports},
  Year                     = {1995},
  Number                   = {6},
  Pages                    = {329 - 365},
  Volume                   = {250},
  Doi                      = {http://dx.doi.org/10.1016/0370-1573(94)00086-I},
  ISSN                     = {0370-1573},
  Url                      = {http://www.sciencedirect.com/science/article/pii/037015739400086I}
}

@article{LeTacon_nphys2011,
   author = {M. Le Tacon and G. Ghiringhelli and J. Chaloupka and M. Moretti Sala and V. Hinkov and M. W. Haverkort and M. Minola and M. Bakr and K. J. Zhou and S. Blanco-Canosa and C. Monney and Y. T. Song and G. L. Sun and C. T. Lin and G. M. De Luca and M. Salluzzo and G. Khaliullin and T. Schmitt and L. Braicovich and B. Keimer},
   doi = {10.1038/nphys2041},
   URL = {https://doi.org/10.1038/nphys2041},
   issn = {17452481},
   issue = {9},
   journal = {Nat. Phys.},
   pages = {725-730},
   publisher = {Nature Publishing Group},
   title = {Intense paramagnon excitations in a large family of high-temperature superconductors},
   volume = {7},
   year = {2011},
}

@article{Dean_nmaterials2013,
   author = {M. P.M. Dean and G. Dellea and R. S. Springell and F. Yakhou-Harris and K. Kummer and N. B. Brookes and X. Liu and Y. J. Sun and J. Strle and T. Schmitt and L. Braicovich and G. Ghiringhelli and I. Božović and J. P. Hill},
   doi = {10.1038/nmat3723},
   issn = {14764660},
   issue = {11},
   journal = {Nat. Mater.},
   pages = {1019-1023},
   publisher = {Nature Publishing Group},
   title = {Persistence of magnetic excitations in {La$_{2-x}$Sr$_x$CuO$_4$} from the undoped insulator to the heavily overdoped non-superconducting metal},
   volume = {12},
   year = {2013},
   note = {\url{https://www.nature.com/articles/nmat3723}},
}

@Article{dean-bisco-PRL,
  Title                    = {High-Energy Magnetic Excitations in the Cuprate Superconductor BiSCCO: Towards a Unified Description of Its Electronic and Magnetic Degrees of Freedom},
  Author                   = {Dean, M. P. M. and James, A. J. A. and Springell, R. S. and Liu, X. and Monney, C. and Zhou, K. J. and Konik, R. M. and Wen, J. S. and Xu, Z. J. and Gu, G. D. and Strocov, V. N. and Schmitt, T. and Hill, J. P.},
  Journal                  = {Phys. Rev. Lett.},
  Year                     = {2013},

  Month                    = {Apr},
  Pages                    = {147001},
  Volume                   = {110},
  Doi                      = {10.1103/PhysRevLett.110.147001},
  Issue                    = {14},
  Numpages                 = {6},
  Publisher                = {American Physical Society},
  Url                      = {http://link.aps.org/doi/10.1103/PhysRevLett.110.147001}
}

@Article{Dean20153,
  Title                    = {Insights into the high temperature superconducting cuprates from resonant inelastic X-ray scattering},
  Author                   = {M.P.M. Dean},
  Journal                  = {Journal of Magnetism and Magnetic Materials},
  Year                     = {2015},
  Note                     = {Pseudogap, Superconductivity and Magnetism},
  Number                   = {0},
  Pages                    = {3 - 13},
  Volume                   = {376},
  Doi                      = {http://dx.doi.org/10.1016/j.jmmm.2014.03.057},
  ISSN                     = {0304-8853},
  Url                      = {http://www.sciencedirect.com/science/article/pii/S0304885314002868}
}

@article{Dellea2017,
   author = {G Dellea and M Minola and A Galdi and D Di Castro and C Aruta and N B Brookes and C J Jia and C Mazzoli and M Moretti Sala and B Moritz and P Orgiani and D G Schlom and A Tebano and G Balestrino and L Braicovich and T P Devereaux and L Maritato and G Ghiringhelli},
   doi = {10.1103/PhysRevB.96.115117},
   issue = {11},
   journal = {Phys. Rev. B},
   month = {9},
   pages = {115117},
   publisher = {American Physical Society},
   title = {Spin and charge excitations in artificial hole- and electron-doped infinite layer cuprate superconductors},
   volume = {96},
   url = {https://link.aps.org/doi/10.1103/PhysRevB.96.115117},
   year = {2017},
}

@article{PengPRB2018,
   title = {Dispersion, Damping, and Intensity of Spin Excitations in the Monolayer ( {{Bi}},{{Pb}} ) 2 ( {{Sr}},{{La}} ) 2 {{CuO}} 6 + {$\delta$} Cuprate Superconductor Family},
   author = {Peng, Y. Y. and Huang, E. W. and Fumagalli, R. and Minola, M. and Wang, Y. and Sun, X. and Ding, Y. and Kummer, K. and Zhou, X. J. and Brookes, N. B. and Moritz, B. and Braicovich, L. and Devereaux, T. P. and Ghiringhelli, G.},
   year = {2018},
   month = {10},
   journal = {Phys. Rev. B},
   volume = {98},
   number = {14},
   pages = {144507},  
   issn = {2469-9950, 2469-9969},
   doi = {10.1103/PhysRevB.98.144507},
   URL = {https://doi.org/10.1103/PhysRevB.98.144507},
   publisher = {American Physical Society},
}

@Article{Baptiste:wm4087,
  Title                    = {{Ca${\sb 2}$CuO${\sb 2}$Cl${\sb 2}$, a redetermination from single-crystal X-ray diffraction data}},
  Author                   = {Baptiste, Beno{\^\i}t and Azuma, Masaki and Yu, Runze and Giura, Paola and d'Astuto, Matteo},
  Journal                  = {IUCrData},
  Year                     = {2018},

  Month                    = {Nov},
  Number                   = {11},
  Pages                    = {x181645},
  Volume                   = {3},
  Doi                      = {10.1107/S2414314618016450},
  Url                      = {https://doi.org/10.1107/S2414314618016450}
}

@article{Braicovich_PRB2010,
   author = {L Braicovich and M Moretti Sala and L J P Ament and V Bisogni and M Minola and G Balestrino and D Di Castro and G M De Luca and M Salluzzo and G Ghiringhelli and J van den Brink},
   doi = {10.1103/PhysRevB.81.174533},
   issue = {17},
   journal = {Phys. Rev. B},
   month = {5},
   pages = {174533},
   publisher = {American Physical Society},
   title = {Momentum and polarization dependence of single-magnon spectral weight for {Cu} {L$_3$}-edge resonant inelastic x-ray scattering from layered cuprates},
   volume = {81},
   url = {https://link.aps.org/doi/10.1103/PhysRevB.81.174533},
   year = {2010},
}

@article{Peng2015,
   author = {Y. Y. Peng and M. Hashimoto and M. Moretti Sala and A. Amorese and N. B. Brookes and G. Dellea and W. S. Lee and M. Minola and T. Schmitt and Y. Yoshida and K. J. Zhou and H. Eisaki and T. P. Devereaux and Z. X. Shen and L. Braicovich and G. Ghiringhelli},
   title = {Magnetic excitations and phonons simultaneously studied by resonant inelastic x-ray scattering in optimally doped ${\mathrm{Bi}}_{1.5}{\mathrm{Pb}}_{0.55}{\mathrm{Sr}}_{1.6}{\mathrm{La}}_{0.4}{\mathrm{CuO}}_{6+\ensuremath{\delta}}$},
   issn = {1550235X},
   journal = {Phys. Rev. B},
   volume = {92},
   issue = {6},
   pages = {064517},
   numpages = {8},
   year = {2015},
   month = {Aug},
  publisher = {American Physical Society},
  doi = {10.1103/PhysRevB.92.064517},
  url = {https://link.aps.org/doi/10.1103/PhysRevB.92.064517}
}

@article{Minola_PRL2015,
  title = {Collective {{Nature}} of {{Spin Excitations}} in {{Superconducting Cuprates Probed}} by {{Resonant Inelastic X-Ray Scattering}}},
  author = {Minola, M. and Dellea, G. and Gretarsson, H. and Peng, Y. Y. and Lu, Y. and Porras, J. and Loew, T. and Yakhou, F. and Brookes, N. B. and Huang, Y. B. and Pelliciari, J. and Schmitt, T. and Ghiringhelli, G. and Keimer, B. and Braicovich, L. and Le Tacon, M.},
  year = {2015},
  month = {may},
  journal = {Phys. Rev. Lett.},
  volume = {114},
  number = {21},
  pages = {217003},
  issn = {0031-9007, 1079-7114},
  doi = {10.1103/PhysRevLett.114.217003},
   URL = {https://doi.org/10.1103/PhysRevLett.114.217003},
}

@article{XiaotingLi_PRL2024,
  title = {Evolution of the Magnetic Excitations in Electron-Doped ${\mathrm{La}}_{2\ensuremath{-}x}{\mathrm{Ce}}_{x}{\mathrm{CuO}}_{4}$},
  author = {Li, X. T. and Tu, S. J. and Chaix, L. and Fawaz, C. and d'Astuto, M. and Li, X. and Yakhou-Harris, F. and Kummer, K. and Brookes, N. B. and Garcia-Fernandez, M. and Zhou, Ke-Jin and Lin, Z. F. and Yuan, J. and Jin, K. and Dean, M. P. M. and Liu, X.},
  journal = {Phys. Rev. Lett.},
  volume = {132},
  issue = {5},
  pages = {056002},
  numpages = {7},
  year = {2024},
  month = {Feb},
  publisher = {American Physical Society},
  doi = {10.1103/PhysRevLett.132.056002},
  url = {https://link.aps.org/doi/10.1103/PhysRevLett.132.056002}
}

@article{Meyers2017,
  title = {Doping dependence of the magnetic excitations in ${\mathrm{La}}_{2\ensuremath{-}x}{\mathrm{Sr}}_{x}{\mathrm{CuO}}_{4}$},
  author = {Meyers, D. and Miao, H. and Walters, A. C. and Bisogni, V. and Springell, R. S. and d'Astuto, M. and Dantz, M. and Pelliciari, J. and Huang, H. Y. and Okamoto, J. and Huang, D. J. and Hill, J. P. and He, X. and Bo{\ifmmode \check{z}\else \v{z}\fi{}}ovi{\ifmmode \acute{c}\else \'{c}\fi{}}, I. and Schmitt, T. and Dean, M. P. M.},
  journal = {Phys. Rev. B},
  volume = {95},
  issue = {7},
  pages = {075139},
  numpages = {7},
  year = {2017},
  month = {Feb},
  publisher = {American Physical Society},
  doi = {10.1103/PhysRevB.95.075139},
  url = {https://link.aps.org/doi/10.1103/PhysRevB.95.075139}
}

@article{Ivashko2017,
   author = {O Ivashko and N E Shaik and X Lu and C G Fatuzzo and M Dantz and P G Freeman and D E McNally and D Destraz and N B Christensen and T Kurosawa and N Momono and M Oda and C E Matt and C Monney and H M Rønnow and T Schmitt and J Chang},
   doi = {10.1103/PhysRevB.95.214508},
   issue = {21},
   journal = {Phys. Rev. B},
   month = {6},
   pages = {214508},
   publisher = {American Physical Society},
   title = {Damped spin excitations in a doped cuprate superconductor with orbital hybridization},
   volume = {95},
   url = {https://link.aps.org/doi/10.1103/PhysRevB.95.214508},
   year = {2017},
}

@article{RobartsPRB2019,
  title = {Anisotropic damping and wave vector dependent susceptibility of the spin fluctuations in ${\mathrm{La}}_{2\ensuremath{-}x}{\mathrm{Sr}}_{x}{\mathrm{CuO}}_{4}$ studied by resonant inelastic x-ray scattering},
  author = {Robarts, H. C. and Barth{\'e}lemy, M. and Kummer, K. and Garc{\'{\i}}a-Fern{\'a}ndez, M. and Li, J. and Nag, A. and Walters, A. C. and Zhou, K. J. and Hayden, S. M.},
  journal = {Phys. Rev. B},
  volume = {100},
  issue = {21},
  pages = {214510},
  numpages = {12},
  year = {2019},
  month = {Dec},
  publisher = {American Physical Society},
  doi = {10.1103/PhysRevB.100.214510},
  url = {https://link.aps.org/doi/10.1103/PhysRevB.100.214510},
}

@Article{Minola2017,
author={Minola, M.
and Lu, Y.
and Peng, Y. Y.
and Dellea, G.
and Gretarsson, H.
and Haverkort, M. W.
and Ding, Y.
and Sun, X.
and Zhou, X. J.
and Peets, D. C.
and Chauviere, L.
and Dosanjh, P.
and Bonn, D. A.
and Liang, R.
and Damascelli, A.
and Dantz, M.
and Lu, X.
and Schmitt, T.
and Braicovich, L.
and Ghiringhelli, G.
and Keimer, B.
and Le Tacon, M.},
title={Crossover from Collective to Incoherent Spin Excitations in Superconducting Cuprates Probed by Detuned Resonant Inelastic X-Ray Scattering},
journal={Phys. Rev. Lett.},
year={2017},
month={Aug},
day={31},
publisher={American Physical Society},
volume={119},
number={9},
pages={097001},
doi={10.1103/PhysRevLett.119.097001},
url={https://doi.org/10.1103/PhysRevLett.119.097001},
}

@Article{Kang2019,
author={Kang, M.
and Pelliciari, J.
and Krockenberger, Y.
and Li, J.
and McNally, D. E.
and Paris, E.
and Liang, R.
and Hardy, W. N.
and Bonn, D. A.
and Yamamoto, H.
and Schmitt, T.
and Comin, R.},
title={Resolving the nature of electronic excitations in resonant inelastic x-ray scattering},
journal={Phys. Rev. B},
year={2019},
month={Jan},
day={02},
publisher={American Physical Society},
volume={99},
number={4},
pages={045105},
doi={10.1103/PhysRevB.99.045105},
url={https://doi.org/10.1103/PhysRevB.99.045105},
}

@article{Fumagalli2019,
  title = {Polarization-resolved Cu ${L}_{3}$-edge resonant inelastic x-ray scattering of orbital and spin excitations in ${\mathrm{NdBa}}_{2}{\mathrm{Cu}}_{3}{\mathrm{O}}_{7\ensuremath{-}\ensuremath{\delta}}$},
  author = {Fumagalli, R. and Braicovich, L. and Minola, M. and Peng, Y. Y. and Kummer, K. and Betto, D. and Rossi, M. and Lefran{\ifmmode \mbox{\c{c}}\else \c{c}\fi{}}ois, E. and Morawe, C. and Salluzzo, M. and Suzuki, H. and Yakhou, F. and Le Tacon, M. and Keimer, B. and Brookes, N. B. and Sala, M. Moretti and Ghiringhelli, G.},
  journal = {Phys. Rev. B},
  volume = {99},
  issue = {13},
  pages = {134517},
  numpages = {12},
  year = {2019},
  month = {Apr},
  publisher = {American Physical Society},
  doi = {10.1103/PhysRevB.99.134517},
  url = {https://link.aps.org/doi/10.1103/PhysRevB.99.134517}

}

@Article{Zhang2022,
author={Zhang, Wenliang
and Agrapidis, Cli{\`o} Efthimia
and Tseng, Yi
and Asmara, Teguh Citra
and Paris, Eugenio
and Strocov, Vladimir N.
and Giannini, Enrico
and Nishimoto, Satoshi
and Wohlfeld, Krzysztof
and Schmitt, Thorsten},
title={Unraveling the nature of spin excitations disentangled from charge contributions in a doped cuprate superconductor},
journal={npj Quantum Materials},
year={2022},
month={Dec},
day={31},
volume={7},
number={1},
pages={123},
issn={2397-4648},
doi={10.1038/s41535-022-00528-5},
url={https://doi.org/10.1038/s41535-022-00528-5},
}

@Article{Tsutsui2016,
author={Tsutsui, Kenji
and Tohyama, Takami},
title={Incident-energy-dependent spectral weight of resonant inelastic x-ray scattering in doped cuprates},
journal={Phys. Rev. B},
year={2016},
month={Aug},
day={26},
publisher={American Physical Society},
volume={94},
number={8},
pages={085144},
doi={10.1103/PhysRevB.94.085144},
url={https://doi.org/10.1103/PhysRevB.94.085144},
}

@article{AmentPRL2009,
  title = {Theoretical Demonstration of How the Dispersion of Magnetic Excitations in Cuprate Compounds can be Determined Using Resonant Inelastic X-Ray Scattering},
  author = {Ament, Luuk J. P. and Ghiringhelli, Giacomo and Sala, Marco Moretti and Braicovich, Lucio and van den Brink, Jeroen},
  journal = {Phys. Rev. Lett.},
  volume = {103},
  issue = {11},
  pages = {117003},
  numpages = {4},
  year = {2009},
  month = {Sep},
  publisher = {American Physical Society},
  doi = {10.1103/PhysRevLett.103.117003},
  url = {https://link.aps.org/doi/10.1103/PhysRevLett.103.117003}
}

@article{MorettiSala_NJP2011,
  title = {Energy and Symmetry of Dd Excitations in Undoped Layered Cuprates Measured by {{Cu}} {{{\emph{L}}}}{\textsubscript{3}} Resonant Inelastic X-Ray Scattering},
  author = {Moretti Sala, M and Bisogni, V and Aruta, C and Balestrino, G and Berger, H and Brookes, N B and Luca, G M De and Di Castro, D and Grioni, M and Guarise, M and Medaglia, P G and Miletto Granozio, F and Minola, M and Perna, P and Radovic, M and Salluzzo, M and Schmitt, T and Zhou, K J and Braicovich, L and Ghiringhelli, G},
  year = {2011},
  month = {apr},
  journal = {New Journal of Physics},
  volume = {13},
  number = {4},
  pages = {043026},
  issn = {1367-2630},
  doi = {10.1088/1367-2630/13/4/043026},
  url={https://dx.doi.org/10.1088/1367-2630/13/4/043026},
}

@Article{Jia2016,
author={Jia, Chunjing
and Wohlfeld, Krzysztof
and Wang, Yao
and Moritz, Brian
and Devereaux, Thomas P.},
title={Using RIXS to Uncover Elementary Charge and Spin Excitations},
journal={Phys. Rev. X},
year={2016},
month={May},
day={13},
publisher={American Physical Society},
volume={6},
number={2},
pages={021020},
doi={10.1103/PhysRevX.6.021020},
url={https://doi.org/10.1103/PhysRevX.6.021020},
}

@article{Arpaia_ncom2023,
  title = {Signature of Quantum Criticality in Cuprates by Charge Density Fluctuations},
  author = {Arpaia, Riccardo and Martinelli, Leonardo and Sala, Marco Moretti and Caprara, Sergio and Nag, Abhishek and Brookes, Nicholas B. and Camisa, Pietro and Li, Qizhi and Gao, Qiang and Zhou, Xingjiang and {Garcia-Fernandez}, Mirian and Zhou, Ke-Jin and Schierle, Enrico and Bauch, Thilo and Peng, Ying Ying and Di Castro, Carlo and Grilli, Marco and Lombardi, Floriana and Braicovich, Lucio and Ghiringhelli, Giacomo},
  year = {2023},
  month = {nov},
  journal = {Nat. Commun.},
  volume = {14},
  number = {1},
  pages = {7198},
  issn = {2041-1723},
  doi = {10.1038/s41467-023-42961-5},
  url = {https://www.nature.com/articles/s41467-023-42961-5},
  urldate = {2025-09-17},
}

@Article{Lamsal2016,
author={Lamsal, Jagat
and Montfrooij, Wouter},
title={Extracting paramagnon excitations from resonant inelastic x-ray scattering experiments},
journal={Phys. Rev. B},
year={2016},
month={Jun},
day={21},
publisher={American Physical Society},
volume={93},
number={21},
pages={214513},
doi={10.1103/PhysRevB.93.214513},
url={https://doi.org/10.1103/PhysRevB.93.214513},
}

@article{MonneyPRB2016,
  title = {Resonant inelastic x-ray scattering study of the spin and charge excitations in the overdoped superconductor ${\mathrm{La}}_{1.77}{\mathrm{Sr}}_{0.23}{\mathrm{CuO}}_{4}$},
  author = {Monney, C. and Schmitt, T. and Matt, C. E. and Mesot, J. and Strocov, V. N. and Lipscombe, O. J. and Hayden, S. M. and Chang, J.},
  journal = {Phys. Rev. B},
  volume = {93},
  issue = {7},
  pages = {075103},
  numpages = {7},
  year = {2016},
  month = {Feb},
  publisher = {American Physical Society},
  doi = {10.1103/PhysRevB.93.075103},
  url = {https://link.aps.org/doi/10.1103/PhysRevB.93.075103}
}

@article{jia2014a,
  title = {Persistent Spin Excitations in Doped Antiferromagnets Revealed by Resonant Inelastic Light Scattering},
  author = {Jia, C. J. and Nowadnick, E. A. and Wohlfeld, K. and Kung, Y. F. and Chen, C.-C. and Johnston, S. and Tohyama, T. and Moritz, B. and Devereaux, T. P.},
  year = 2014,
  month = feb,
  journal = {Nat. Commun.},
  volume = {5},
  number = {1},
  pages = {3314},
  publisher = {Nature Publishing Group},
  issn = {2041-1723},
  doi = {10.1038/ncomms4314},
  URL = {https://doi.org/10.1038/ncomms4314},
  urldate = {2026-04-09},
  copyright = {2014 Springer Nature Limited},
  langid = {english}
}

@article{stepanovQuantumSpinFluctuations2018,
  title = {Quantum Spin Fluctuations and Evolution of Electronic Structure in Cuprates},
  author = {Stepanov, Evgeny A. and Peters, Lars and Krivenko, Igor S. and Lichtenstein, Alexander I. and Katsnelson, Mikhail I. and Rubtsov, Alexey N.},
  year = 2018,
  month = oct,
  journal = {npj Quantum Materials},
  volume = {3},
  number = {1},
  pages = {54},
  publisher = {Nature Publishing Group},
  issn = {2397-4648},
  doi = {10.1038/s41535-018-0128-x},
  URL = {https://doi.org/10.1038/s41535-018-0128-x},
  urldate = {2026-04-09},
  copyright = {2018 The Author(s)},
  langid = {english}
}

@article{giannozziQUANTUMESPRESSOModular2009,
  title = {{{QUANTUM ESPRESSO}}: A Modular and Open-Source Software Project for Quantum Simulations of Materials},
  shorttitle = {{{QUANTUM ESPRESSO}}},
  author = {Giannozzi, Paolo and Baroni, Stefano and Bonini, Nicola and Calandra, Matteo and Car, Roberto and Cavazzoni, Carlo and Ceresoli, Davide and Chiarotti, Guido L. and Cococcioni, Matteo and Dabo, Ismaila and Dal Corso, Andrea and {de Gironcoli}, Stefano and Fabris, Stefano and Fratesi, Guido and Gebauer, Ralph and Gerstmann, Uwe and Gougoussis, Christos and Kokalj, Anton and Lazzeri, Michele and {Martin-Samos}, Layla and Marzari, Nicola and Mauri, Francesco and Mazzarello, Riccardo and Paolini, Stefano and Pasquarello, Alfredo and Paulatto, Lorenzo and Sbraccia, Carlo and Scandolo, Sandro and Sclauzero, Gabriele and Seitsonen, Ari P. and Smogunov, Alexander and Umari, Paolo and Wentzcovitch, Renata M.},
  year = 2009,
  month = sep,
  journal = {Journal of Physics. Condensed Matter: An Institute of Physics Journal},
  volume = {21},
  number = {39},
  pages = {395502},
  issn = {1361-648X},
  doi = {10.1088/0953-8984/21/39/395502},
  URL = {https://doi.org/10.1088/0953-8984/21/39/395502},
  langid = {english},
  pmid = {21832390}
}

@article{perdewGeneralizedGradientApproximation1996,
  title = {Generalized {{Gradient Approximation Made Simple}}},
  author = {Perdew, John P. and Burke, Kieron and Ernzerhof, Matthias},
  year = 1996,
  month = oct,
  journal = {Phys. Rev. Lett.},
  volume = {77},
  number = {18},
  pages = {3865--3868},
  publisher = {American Physical Society},
  doi = {10.1103/PhysRevLett.77.3865},
  URL = {https://doi.org/10.1103/PhysRevLett.77.3865},
  urldate = {2023-05-09}
}

@article{hamannOptimizedNormconservingVanderbilt2013a,
  title = {Optimized Norm-Conserving {{Vanderbilt}} Pseudopotentials},
  author = {Hamann, D. R.},
  year = 2013,
  month = aug,
  journal = {Phys. Rev. B},
  volume = {88},
  number = {8},
  pages = {085117},
  publisher = {American Physical Society},
  doi = {10.1103/PhysRevB.88.085117},
  URL = {https://doi.org/10.1103/PhysRevB.88.085117},
  urldate = {2026-04-09}
}

@article{marzariMaximallyLocalizedGeneralized1997,
  title = {Maximally Localized Generalized {{Wannier}} Functions for Composite Energy Bands},
  author = {Marzari, Nicola and Vanderbilt, David},
  year = 1997,
  month = nov,
  journal = {Phys. Rev. B},
  volume = {56},
  number = {20},
  pages = {12847--12865},
  publisher = {American Physical Society},
  doi = {10.1103/PhysRevB.56.12847},
  URL = {ttps://doi.org/10.1103/PhysRevB.56.12847},
  urldate = {2021-12-02}
}

@article{souzaMaximallyLocalizedWannier2001,
  title = {Maximally Localized {{Wannier}} Functions for Entangled Energy Bands},
  author = {Souza, Ivo and Marzari, Nicola and Vanderbilt, David},
  year = 2001,
  month = dec,
  journal = {Phys. Rev. B},
  volume = {65},
  number = {3},
  pages = {035109},
  publisher = {American Physical Society},
  doi = {10.1103/PhysRevB.65.035109},
  URL = {https://doi.org/10.1103/PhysRevB.65.035109},
  urldate = {2021-12-02}
}

@article{wietekXDiagExactDiagonalization2025,
	title = {XDiag: Exact diagonalization for quantum many-body systems},
	pages = {70},
	author = {Wietek, Alexander and Staszewski, Luke and Ulaga, Martin and Ebert, Paul L. and Karlsson, Hannes and Sarkar, Siddhartha and Shackleton, Leyna and Sinha, Aritra and Diogo Soares, Rafael},
	journal = {SciPost Phys. Codebases},
	year = {2026},
	publisher = {SciPost},
	doi = {10.21468/SciPostPhysCodeb.70},
	url = {https://scipost.org/10.21468/SciPostPhysCodeb.70}
}

@article{tohyamaAsymmetryElectronicStates2004,
  title = {Asymmetry of the Electronic States in Hole- and Electron-Doped Cuprates: {{Exact}} Diagonalization Study of the \$t\textbackslash text\textbraceleft\textbackslash ensuremath\textbraceleft -\textbraceright\textbraceright\textbraceleft t\textbraceright\textasciicircum\textbraceleft\textbackslash ensuremath\textbraceleft '\textbraceright\textbraceright\textbackslash text\textbraceleft\textbackslash ensuremath\textbraceleft -\textbraceright\textbraceright\textbraceleft t\textbraceright\textasciicircum\textbraceleft\textbackslash ensuremath\textbraceleft ''\textbraceright\textbraceright\textbackslash text\textbraceleft\textbackslash ensuremath\textbraceleft -\textbraceright\textbraceright{{J}}\$ Model},
  shorttitle = {Asymmetry of the Electronic States in Hole- and Electron-Doped Cuprates},
  author = {Tohyama, T.},
  year = 2004,
  month = nov,
  journal = {Phys. Rev. B},
  volume = {70},
  number = {17},
  pages = {174517},
  publisher = {American Physical Society},
  doi = {10.1103/PhysRevB.70.174517},
  URL = {https://doi.org/10.1103/PhysRevB.70.174517},
  urldate = {2025-12-05}
}

@article{poilblancTwistedBoundaryConditions1991,
  title = {Twisted Boundary Conditions in Cluster Calculations of the Optical Conductivity in Two-Dimensional Lattice Models},
  author = {Poilblanc, Didier},
  year = 1991,
  month = nov,
  journal = {Phys. Rev. B},
  volume = {44},
  number = {17},
  pages = {9562--9581},
  publisher = {American Physical Society},
  doi = {10.1103/PhysRevB.44.9562},
  URL = {https://doi.org/10.1103/PhysRevB.44.9562},
  urldate = {2026-01-19}
}

@article{bettsImprovedFinitelatticeEstimates1999,
  title = {Improved Finite-Lattice Estimates of the Properties of Two Quantum Spin Models on the Infinite Square Lattice},
  author = {Betts, D D and Lin, H Q and Flynn, J S},
  year = 1999,
  month = sep,
  journal = {Canadian Journal of Physics},
  volume = {77},
  number = {5},
  pages = {353--369},
  publisher = {NRC Research Press},
  issn = {0008-4204},
  doi = {10.1139/p99-041},
  URL = { https://doi.org/10.1139/p99-041},
  urldate = {2025-11-13}
}

@article{bettsImprovedFinitelatticeMethod1996,
  title = {Improved Finite-Lattice Method for Estimating the Zero-Temperature Properties of Two-Dimensional Lattice Models},
  author = {Betts, D. D. and Masui, S. and Vats, N. and Stewart, G. E.},
  year = 1996,
  month = jan,
  journal = {Canadian Journal of Physics},
  volume = {74},
  number = {1-2},
  pages = {54--64},
  publisher = {NRC Research Press},
  issn = {0008-4204},
  doi = {10.1139/p96-010},
  URL = {https://doi.org/10.1139/p96-010},
  urldate = {2025-11-13}
}

@article{betto2021,
  title = {Multiple-Magnon Excitations Shape the Spin Spectrum of Cuprate Parent Compounds},
  author = {Betto, Davide and Fumagalli, Roberto and Martinelli, Leonardo and Rossi, Matteo and Piombo, Riccardo and Yoshimi, Kazuyoshi and Di Castro, Daniele and Di Gennaro, Emiliano and Sambri, Alessia and Bonn, Doug and Sawatzky, George A. and Braicovich, Lucio and Brookes, Nicholas B. and Lorenzana, Jos{\'e} and Ghiringhelli, Giacomo},
  year = 2021,
  month = apr,
  journal = {Phys. Rev. B},
  volume = {103},
  number = {14},
  pages = {L140409},
  issn = {2469-9950, 2469-9969},
  doi = {10.1103/PhysRevB.103.L140409},
  URL = {https://doi.org/10.1103/PhysRevB.103.L140409},
  urldate = {2025-09-17},
  langid = {english},
}

@article{linTwistaveragedBoundaryConditions2001,
  title = {Twist-Averaged Boundary Conditions in Continuum Quantum {{Monte Carlo}} Algorithms},
  author = {Lin, C. and Zong, F. H. and Ceperley, D. M.},
  year = 2001,
  month = jun,
  journal = {Phys. Rev. E},
  volume = {64},
  number = {1},
  pages = {016702},
  publisher = {American Physical Society},
  doi = {10.1103/PhysRevE.64.016702},
  url = {https://doi.org/10.1103/PhysRevE.64.016702},
  urldate = {2026-02-04}
}

@Unpublished{ChaficRamanCCOC,
  author      = {Fawaz, Chafic and Gao, Yingzheng and Laveder, Luca and Menon, Lorenzo and Moulding, Owen and Heid, Rolf and Lebert, Blair W and Bellin, Christophe and Beneut, Keevin and Santos-Cottin, David and Yamada, Ikuya and Okazaki, Yuichi and Yamamoto, Hajime and Azuma, Masaki and M{\'e}asson, M.-A. and d'Astuto, Matteo},
  title       = {{Raman scattering in cuprate oxychlorides high-temperature superconductors single-crystals}},
  note        = {working paper or preprint},
  month       = Sep,
  year        = {2026},
  hal_id      = {hal-05742802},
  hal_version = {v1},
  url         = {https://hal.science/hal-05742802},
}

@article{Daghofer_2008,
  title = {Absence of Hole Confinement in Transition-Metal Oxides with Orbital Degeneracy},
  author = {Daghofer, Maria and Wohlfeld, Krzysztof and Ole\ifmmode \acute{s}\else \'{s}\fi{}, Andrzej M. and Arrigoni, Enrico and Horsch, Peter},
  journal = {Phys. Rev. Lett.},
  volume = {100},
  issue = {6},
  pages = {066403},
  numpages = {4},
  year = {2008},
  month = {Feb},
  publisher = {American Physical Society},
  doi = {10.1103/PhysRevLett.100.066403},
  url = {https://link.aps.org/doi/10.1103/PhysRevLett.100.066403}
}

@article{huang2018stripe,
  title={Stripe order from the perspective of the Hubbard model},
  author={Huang, Edwin W and Mendl, Christian B and Jiang, Hong-Chen and Moritz, Brian and Devereaux, Thomas P},
  journal={npj Quantum Materials},
  volume={3},
  number={1},
  pages={22},
  year={2018},
  publisher={Nature Publishing Group UK London},
  DOI = {10.1038/s41535-018-0097-0},
  URL = {https://doi.org/10.1038/s41535-018-0097-0},
}

@article{xu2022stripes,
  title={Stripes and spin-density waves in the doped two-dimensional Hubbard model: Ground state phase diagram},
  author={Xu, Hao and Shi, Hao and Vitali, Ettore and Qin, Mingpu and Zhang, Shiwei},
  journal={Phys. Rev. Res.},
  volume={4},
  number={1},
  pages={013239},
  numpages = {10},
  year={2022},
  publisher = {American Physical Society},
  doi = {10.1103/PhysRevResearch.4.013239},
  url = {https://link.aps.org/doi/10.1103/PhysRevResearch.4.013239}
}

@article{xu2024coexistence,
  title={Coexistence of superconductivity with partially filled stripes in the Hubbard model},
  author={Xu, Hao and Chung, Chia-Min and Qin, Mingpu and Schollw{\"o}ck, Ulrich and White, Steven R and Zhang, Shiwei},
  journal={Science},
  volume={384},
  number={6696},
  pages={eadh7691},
  year={2024},
  publisher={American Association for the Advancement of Science},
  DOI = {10.1126/science.adh7691},
  URL = {https://doi.org/10.1126/science.adh7691},
}

@article{zheng2017stripe,
  title={Stripe order in the underdoped region of the two-dimensional Hubbard model},
  author={Zheng, Bo-Xiao and Chung, Chia-Min and Corboz, Philippe and Ehlers, Georg and Qin, Ming-Pu and Noack, Reinhard M and Shi, Hao and White, Steven R and Zhang, Shiwei and Chan, Garnet Kin-Lic},
  journal={Science},
  volume={358},
  number={6367},
  pages={1155--1160},
  year={2017},
  publisher={American Association for the Advancement of Science},
  doi = {10.1126/science.aam7127},
  URL = {https://doi.org/10.1126/science.aam7127},
}

@article{zaanen1985band,
  title = {Band gaps and electronic structure of transition-metal compounds},
  author = {Zaanen, J. and Sawatzky, G. A. and Allen, J. W.},
  journal = {Phys. Rev. Lett.},
  volume = {55},
  issue = {4},
  pages = {418--421},
  numpages = {0},
  year = {1985},
  month = {Jul},
  publisher = {American Physical Society},
  doi = {10.1103/PhysRevLett.55.418},
  url = {https://link.aps.org/doi/10.1103/PhysRevLett.55.418}
}

@article{emery1987theory,
  title = {Theory of high-${\mathrm{T}}_{\mathrm{c}}$ superconductivity in oxides},
  author = {Emery, V. J.},
  journal = {Phys. Rev. Lett.},
  volume = {58},
  issue = {26},
  pages = {2794--2797},
  numpages = {0},
  year = {1987},
  month = {Jun},
  publisher = {American Physical Society},
  doi = {10.1103/PhysRevLett.58.2794},
  url = {https://link.aps.org/doi/10.1103/PhysRevLett.58.2794}
}

@article{eliashberg1960,
Author = {Eliashberg, G. M.},
Title = {Interactions between Electrons and Lattice Vibrations in a Superconductor},
Journal = {Sov. Phys. JETP},
Year = {1960},
Volume = {11},
Number = {3},
Pages = {696--702},
ISSN = {0038-5646},
}

@article{dahm2009,
  title={Strength of the spin-fluctuation-mediated pairing interaction in a high-temperature superconductor},
  author={Dahm, Thomas and Hinkov, V and Borisenko, SV and Kordyuk, AA and Zabolotnyy, VB and Fink, J and B{\"u}chner, B and Scalapino, DJ and Hanke, W and Keimer, B},
  journal={Nat. Phys.},
  volume={5},
  number={3},
  pages={217--221},
  year={2009},
  publisher={Nature Publishing Group UK London},
  doi={https://doi.org/10.1038/nphys1180},
  URL = {https://doi.org/10.1038/nphys1180},
}

@article{simkovich2024,
author = {Fedor Šimkovic  and Riccardo Rossi  and Antoine Georges  and Michel Ferrero },
title = {Origin and fate of the pseudogap in the doped Hubbard model},
journal = {Science},
volume = {385},
number = {6715},
pages = {eade9194},
year = {2024},
doi = {10.1126/science.ade9194},
URL = {https://www.science.org/doi/abs/10.1126/science.ade9194},
}

@article{dong2022,
  title={Quantifying the role of antiferromagnetic fluctuations in the superconductivity of the doped Hubbard model},
  author={Dong, Xinyang and Gull, Emanuel and Millis, Andrew J},
  journal={Nat. Phys.},
  volume={18},
  number={11},
  pages={1293--1296},
  year={2022},
  publisher={Nature Publishing Group UK London},
  doi={https://doi.org/10.1038/s41567-022-01710-z}
}

@article{bacq2025_prx,
  title = {Toward an Ab Initio Theory of High-Temperature Superconductors: A Study of Multilayer Cuprates},
  author = {Bacq-Labreuil, Benjamin and Lacasse, Benjamin and Tremblay, A.-M. S. and S\'en\'echal, David and Haule, Kristjan},
  journal = {Phys. Rev. X},
  volume = {15},
  issue = {2},
  pages = {021071},
  numpages = {44},
  year = {2025},
  month = {May},
  publisher = {American Physical Society},
  doi = {10.1103/PhysRevX.15.021071},
  url = {https://link.aps.org/doi/10.1103/PhysRevX.15.021071}
}

@article{simkovic2022,
  title = {Two-dimensional Hubbard model at finite temperature: Weak, strong, and long correlation regimes},
  author = {Fedor {\ifmmode \check{S}\else \v{S}\fi{}}imkovic IV and Rossi, Riccardo and Ferrero, Michel},
  journal = {Phys. Rev. Res.},
  volume = {4},
  issue = {4},
  pages = {043201},
  numpages = {21},
  year = {2022},
  month = {Dec},
  publisher = {American Physical Society},
  doi = {10.1103/PhysRevResearch.4.043201},
  url = {https://link.aps.org/doi/10.1103/PhysRevResearch.4.043201}
}

@article{PhysRevB.90.085102,
  title = {Efficient real-frequency solver for dynamical mean-field theory},
  author = {Lu, Y. and H\"oppner, M. and Gunnarsson, O. and Haverkort, M. W.},
  journal = {Phys. Rev. B},
  volume = {90},
  issue = {8},
  pages = {085102},
  numpages = {18},
  year = {2014},
  month = {Aug},
  publisher = {American Physical Society},
  doi = {10.1103/PhysRevB.90.085102},
  url = {https://link.aps.org/doi/10.1103/PhysRevB.90.085102}
}

@article{Haverkort_2014,
doi = {10.1209/0295-5075/108/57004},
url = {https://doi.org/10.1209/0295-5075/108/57004},
year = {2014},
month = {dec},
publisher = {EDP Sciences, IOP Publishing and Società Italiana di Fisica},
volume = {108},
number = {5},
pages = {57004},
author = {Haverkort, M. W. and Sangiovanni, G. and Hansmann, P. and Toschi, A. and Lu, Y. and Macke, S.},
title = {Bands, resonances, edge singularities and excitons in core level spectroscopy investigated within the dynamical mean-field theory},
journal = {Europhysics Letters}
}

@article{PhysRevB.85.165113,
  title = {Multiplet ligand-field theory using Wannier orbitals},
  author = {Haverkort, M. W. and Zwierzycki, M. and Andersen, O. K.},
  journal = {Phys. Rev. B},
  volume = {85},
  issue = {16},
  pages = {165113},
  numpages = {20},
  year = {2012},
  month = {Apr},
  publisher = {American Physical Society},
  doi = {10.1103/PhysRevB.85.165113},
  url = {https://link.aps.org/doi/10.1103/PhysRevB.85.165113}
}

@Article{Fleury_Loudon_1968,
  author    = {P A Fleury and R Loudon},
  title     = {Scattering of Light by One- and Two-Magnon Excitations},
  journal   = {Phys. Rev.},
  year      = {1968},
  volume    = {166},
  pages     = {514-530},
  month     = {2},
  doi       = {10.1103/PhysRev.166.514},
  issue     = {2},
  publisher = {American Physical Society},
  url       = {https://doi.org/10.1103/PhysRev.166.514},
}

@article{devereauxILS2007,
  title = {Inelastic Light Scattering from Correlated Electrons},
  author = {Devereaux, Thomas P. and Hackl, Rudi},
  year = 2007,
  month = jan,
  journal = {Reviews of Modern Physics},
  volume = {79},
  number = {1},
  pages = {175--233},
  issn = {0034-6861, 1539-0756},
  doi = {10.1103/RevModPhys.79.175},
  url = {https://link.aps.org/doi/10.1103/RevModPhys.79.175},
  urldate = {2025-09-24},
  copyright = {http://link.aps.org/licenses/aps-default-license},
  langid = {english},
}

@article{Bisogni_PRB2012,
  title = {Bimagnon studies in cuprates with resonant inelastic x-ray scattering at the O $K$ edge. I. Assessment on La${}_{2}$CuO${}_{4}$ and comparison with the excitation at Cu ${L}_{3}$ and Cu $K$ edges},
  author = {Bisogni, V. and Simonelli, L. and Ament, L. J. P. and Forte, F. and Moretti Sala, M. and Minola, M. and Huotari, S. and van den Brink, J. and Ghiringhelli, G. and Brookes, N. B. and Braicovich, L.},
  journal = {Phys. Rev. B},
  volume = {85},
  issue = {21},
  pages = {214527},
  numpages = {9},
  year = {2012},
  month = {Jun},
  publisher = {American Physical Society},
  doi = {10.1103/PhysRevB.85.214527},
  url = {https://link.aps.org/doi/10.1103/PhysRevB.85.214527}
}

@article{forte2008,
  title = {Magnetic Excitations in {{La}} 2 {{Cu O}} 4 Probed by Indirect Resonant Inelastic X-Ray Scattering},
  author = {Forte, Filomena and Ament, Luuk J. P. and Van Den Brink, Jeroen},
  year = 2008,
  month = apr,
  journal = {Phys. Rev. B},
  volume = {77},
  number = {13},
  pages = {134428},
  issn = {1098-0121, 1550-235X},
  doi = {10.1103/PhysRevB.77.134428},
  url = {https://link.aps.org/doi/10.1103/PhysRevB.77.134428},
  urldate = {2026-09-17},
  copyright = {http://link.aps.org/licenses/aps-default-license},
  langid = {english},
}

@Article{Brookes2018,
  author   = {N. B. Brookes and F. Yakhou-Harris and K. Kummer and A. Fondacaro and J.C. Cezar and D. Betto and E. Velez-Fort and A. Amorese and G. Ghiringhelli and L. Braicovich and R. Barrett and G. Berruyer and F. Cianciosi and L. Eybert and P. Marion and P. {van der Linden} and L. Zhang},
  title    = {The beamline ID32 at the ESRF for soft X-ray high energy resolution resonant inelastic X-ray scattering and polarisation dependent X-ray absorption spectroscopy},
  journal  = {Nuclear Instruments and Methods in Physics Research Section A: Accelerators, Spectrometers, Detectors and Associated Equipment},
  year     = {2018},
  volume   = {903},
  pages    = {175-192},
  issn     = {0168-9002},
  doi      = {https://doi.org/10.1016/j.nima.2018.07.001},
  url      = {https://www.sciencedirect.com/science/article/pii/S0168900218308234},
}

@article{braicovichMagneticExcitationsPhase2010,
  title = {Magnetic {{Excitations}} and {{Phase Separation}} in the {{Underdoped La}} 2 - x {{Sr}} x {{CuO}} 4 {{Superconductor Measured}} by {{Resonant Inelastic X-Ray Scattering}}},
  author = {Braicovich, L. and Van Den Brink, J. and Bisogni, V. and Sala, M. Moretti and Ament, L. J. P. and Brookes, N. B. and De Luca, G. M. and Salluzzo, M. and Schmitt, T. and Strocov, V. N. and Ghiringhelli, G.},
  year = 2010,
  month = feb,
  journal = {Phys. Rev. Lett.},
  volume = {104},
  number = {7},
  pages = {077002},
  issn = {0031-9007, 1079-7114},
  doi = {10.1103/PhysRevLett.104.077002},
  url= {https://link.aps.org/doi/10.1103/PhysRevLett.104.077002},
  urldate = {2026-09-17},
  copyright = {http://link.aps.org/licenses/aps-default-license},
  langid = {english},
}

@article{singhBimagnonDispersionLa2CuO42025,
  title = {Bimagnon Dispersion of {{La2CuO4}} Probed by Resonant Inelastic {{X-ray}} Scattering},
  author = {Singh, A. and Huang, H. Y. and Tsutsui, K. and Tohyama, T. and Komiya, S. and Okamoto, J. and Chen, C. T. and Fujimori, A. and Huang, D. J.},
  year = 2025,
  month = oct,
  journal = {Scientific Reports},
  volume = {15},
  number = {1},
  pages = {34183},
  issn = {2045-2322},
  doi = {10.1038/s41598-025-15435-5},
  url = {https://www.nature.com/articles/s41598-025-15435-5},
  urldate = {2026-07-06},
  langid = {english},
}

@article{AmentRMP2011,
  title = {Resonant inelastic x-ray scattering studies of elementary excitations},
  author = {Ament, Luuk J. P. and van Veenendaal, Michel and Devereaux, Thomas P. and Hill, John P. and van den Brink, Jeroen},
  journal = {Rev. Mod. Phys.},
  volume = {83},
  issue = {2},
  pages = {705--767},
  numpages = {0},
  year = {2011},
  month = {Jun},
  publisher = {American Physical Society},
  doi = {10.1103/RevModPhys.83.705},
  url = {https://link.aps.org/doi/10.1103/RevModPhys.83.705}
}

@Article{muschler2010,
  author    = {Muschler, B. and Prestel, W. and Tassini, L. and Hackl, R. and Lambacher, M. and Erb, A. and Komiya, Seiki and Ando, Yoichi and Peets, D.C. and Hardy, W.N. and Liang, R. and Bonn, D.A.},
  title     = {Electron interactions and charge ordering in CuO 2 compounds},
  journal   = {Eur. Phys. J. Spec. Top.},
  year      = {2010},
  volume    = {188},
  number    = {1},
  pages     = {131--152},
  month     = Oct,
  issn      = {1951-6401},
  doi       = {10.1140/epjst/e2010-01302-4},
  publisher = {Springer Science and Business Media LLC},
  url       = {https://doi.org/10.1140/epjst/e2010-01302-4},
}

@article{Chao1977tJ,
  author  = {Chao, K. A. and Spa{\l}ek, J. and Ole{\'s}, A. M.},
  title   = {Canonical Perturbation Expansion of the Hubbard Model},
  journal = {Journal of Physics C: Solid State Physics},
  volume  = {10},
  number  = {10},
  pages   = {L271--L276},
  issn = {0022-3719},
  year    = {1977},
  doi     = {10.1088/0022-3719/10/10/002},
  URL     = {https://doi.org/10.1088/0022-3719/10/10/002},
}

@article{chaixRIXS2018,
  title = {Resonant Inelastic X-Ray Scattering Studies of Magnons and Bimagnons in the Lightly Doped Cuprate {{La}} 2 - x {{Sr}} x {{CuO}} 4},
  author = {Chaix, L. and Huang, E. W. and Gerber, S. and Lu, X. and Jia, C. and Huang, Y. and McNally, D. E. and Wang, Y. and Vernay, F. H. and Keren, A. and Shi, M. and Moritz, B. and Shen, Z.-X. and Schmitt, T. and Devereaux, T. P. and Lee, W.-S.},
  year = 2018,
  month = apr,
  journal = {Phys. Rev. B},
  volume = {97},
  number = {15},
  pages = {155144},
  issn = {2469-9950, 2469-9969},
  doi = {10.1103/PhysRevB.97.155144},
  url = {https://link.aps.org/doi/10.1103/PhysRevB.97.155144},
  urldate = {2026-09-28},
  langid = {english}
}

@article{braicovich2014,
	title = {The Simultaneous Measurement of Energy and Linear Polarization of the Scattered Radiation in Resonant Inelastic Soft X-Ray Scattering},
	author = {Braicovich, L. and Minola, M. and Dellea, G. and Le Tacon, M. and Moretti Sala, M. and Morawe, C. and Peffen, J.-Ch. and Supruangnet, R. and Yakhou, F. and Ghiringhelli, G. and Brookes, N. B.},
	year = 2014,
	month = nov,
	journal = {Review of Scientific Instruments},
	volume = {85},
	number = {11},
	pages = {115104},
	issn = {0034-6748, 1089-7623},
	doi = {10.1063/1.4900959},
	url = {https://pubs.aip.org/rsi/article/85/11/115104/356333/The-simultaneous-measurement-of-energy-and-linear},
}

@article{LeTaconPRB2013,
	title = {Dispersive spin excitations in highly overdoped cuprates revealed by resonant inelastic x-ray scattering},
	author = {Le Tacon, M. and Minola, M. and Peets, D. C. and Moretti Sala, M. and Blanco-Canosa, S. and Hinkov, V. and Liang, R. and Bonn, D. A. and Hardy, W. N. and Lin, C. T. and Schmitt, T. and Braicovich, L. and Ghiringhelli, G. and Keimer, B.},
	journal = {Phys. Rev. B},
	volume = {88},
	issue = {2},
	pages = {020501(R)},
	numpages = {5},
	year = {2013},
	month = {Jul},
	publisher = {American Physical Society},
	doi = {10.1103/PhysRevB.88.020501},
	url = {https://link.aps.org/doi/10.1103/PhysRevB.88.020501}
}

@misc{esrf_experiment_5427,
  note      = {Experiment dataset from HC5427, European Synchrotron Radiation Facility. DOI: \url{https://doi.org/10.15151/esrf-dc-2527566226}},
}

@misc{esrf_experiment_5226,
  note      = {Experiment dataset from HC5226, European Synchrotron Radiation Facility. DOI: \url{https://doi.org/10.15151/ESRF-ES-1176725716}},
}

@book{Auerbach1994,
  author    = {Auerbach, Assa},
  title     = {Interacting Electrons and Quantum Magnetism},
  publisher = {Springer},
  year      = {1994}
}

\appendix

\section{Selection rules}\label{app:Polarization_dependency}
In this work, we took advantage of polarization analysis, along with geometrical factors, to separate the spin-flip (single paramagnon) from the non-spin-flip (including the bimagnon and other possible excitations) in the magnetic board excitation.

We adopt the calculation made by Li \textit{et al.} \cite{XiaotingLi_PRL2024} based on the RIXS cross-section calculation using a single ion model introduced earlier \cite{AmentPRL2009, MorettiSala_NJP2011}. 
Considering a $3d_{x^2 - y^2} \rightleftarrows  2p_{3/2} $ transition, the scattering amplitude is given by \cite{XiaotingLi_PRL2024}: 

\begin{equation}
\begin{aligned}
I^{\downarrow (\uparrow)} (\theta_s = 90^\circ, \epsilon, \epsilon')
\propto \Bigg|
\sum_{m= \pm \frac{3}{2}, \pm \frac{1}{2}}
&\bra{d^{\downarrow (\uparrow)}_{x^2 - y^2}}
\, \epsilon'^{\dag} \cdot \mathbf{r} \,
\ket{ p_{3/2,m} } \\
&\times
\bra{p_{3/2,m}}
\, \epsilon \cdot \mathbf{r} \,
\ket{d^\downarrow_{x^2 - y^2}}
\Bigg|^2
\end{aligned}
\label{Kramers_cross}
\end{equation}

where $\theta_s$ is the angle between the $ab-$ plane and the spin direction,  $\epsilon (\epsilon')$ and $\mathbf{r}$ are the polarization and the position operator. The calculations of this intensity are thus given in Table~\ref{pol_table1}.
\begin{table}[!ht]
    \centering
    \begin{tabular}{c|c|cc}
     \hline \hline
                                    &\diagbox{Out}{In}& $\sigma_{in}$ & $\pi_{in}$ \\
     \hline \hline
      Spin-flip                     & $\sigma_{out}$    & 0 & $\sin^2\alpha \sin^2\theta_s $ \\
      $I_{\Delta S = 1}(\theta_s)$  & $\pi_{out}$       & $\sin^2 \beta \sin^2 \theta_s$ & 0\\
      \hline
      Non-spin-flip                 & $\sigma_{out}$   & 4 & $\sin^2 \alpha \cos^2 \theta_s$\\
      $I_{\Delta S = 0}(\theta_s)$  & $\pi_{out}$      & $\sin^2 \beta \cos^2 \theta_s$ & 4 $\sin^2 \alpha \sin^2 \beta$\\
      \hline \hline
    \end{tabular}
    \caption{Selection rules for the spin-flip and non-spin-flip signal based on the incident and scattering polarizations, for details see Ref.~\onlinecite{XiaotingLi_PRL2024}.}
    \label{pol_table1}
\end{table}
The angles $\alpha$ and $\beta$ are the incident and scattering angles, as noted in Fig.~\ref{experiments_geometry}. In the case of \CCOC\, the spins are in the $ab$-plane as found by single crystal neutron diffraction \cite{vaknin_prb}, therefore $\theta_s = 90^\circ$. Note that the in-plane direction of the spin is not important since it vanishes in the calculation. Therefore, we can rewrite the terms as in Table~\ref{main:pol_table2}.

These results imply that the spin-flip is always seen in the crossed-polarized channel, while the non-spin-flip is in the parallel-polarized channel.

Note that the distinction between the spin-flip and non-spin-flip channels can be extended to other $3d \rightleftarrows  2p $ transitions and are not exclusively limited to magnetic excitations. For example, we can have a spin-flip process within the \textit{dd}-orbital. Similarly, the non-spin-flip channel also includes the elastic line and charge excitations \cite{MorettiSala_NJP2011, Jia2016}.

We have to emphasize that this model is only used qualitatively.

\section{Data analysis}\label{data_analysis}

\begin{figure}[!htb]
    \centering
    \includegraphics[width=0.4\textwidth]{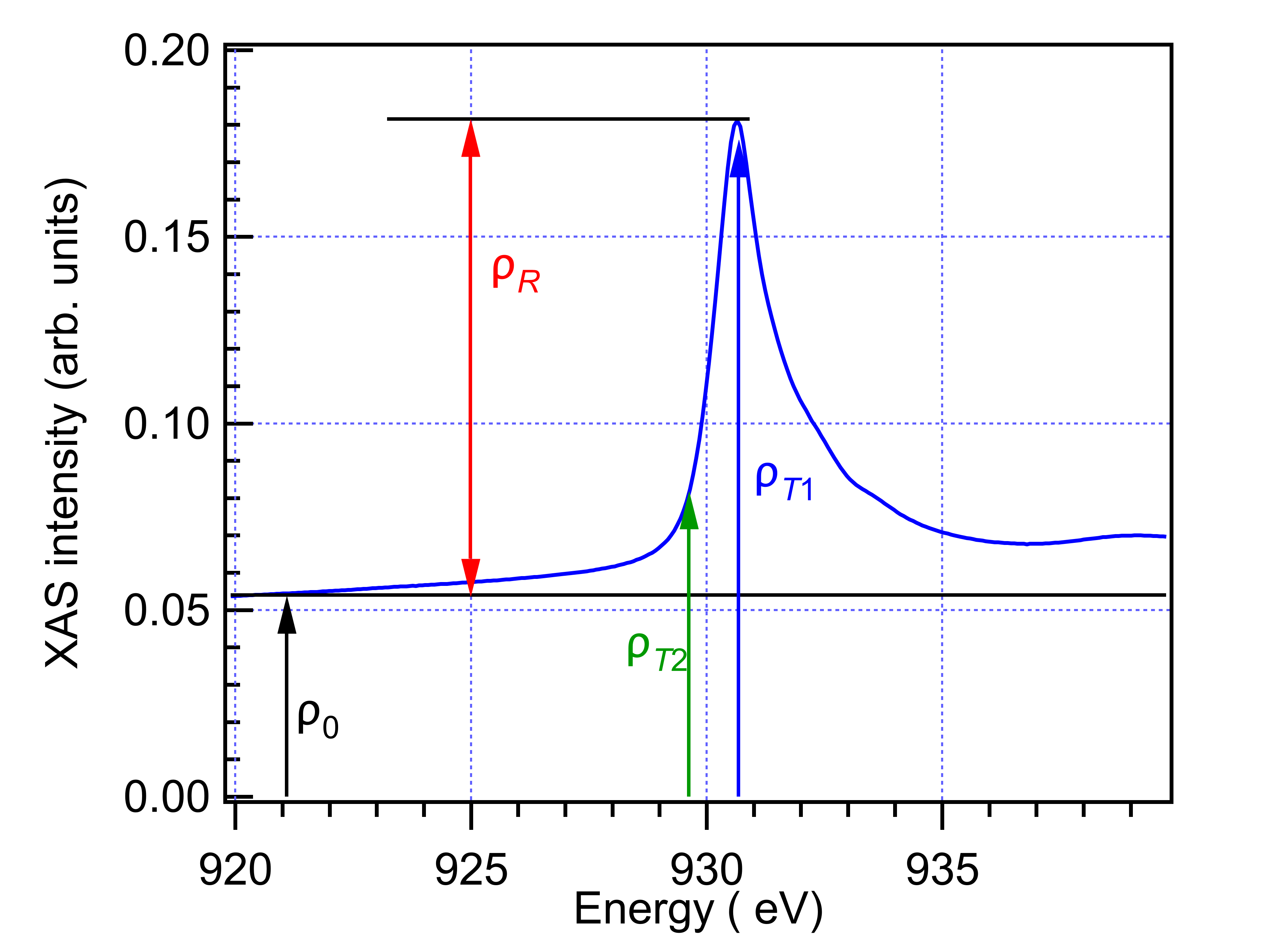}
    \caption{Example of X-ray absorption spectra of maximally doped \NaCCOC\ measured in this work, where we indicate the energy position and value for $\rho_0$, $\rho_R$, $\rho_{T1}$ and $\rho_{T2}$ as discussed in the text.}
    \label{XAS_absorption}
\end{figure}

\subsection{Self-absorption correction}\label{Self_absorption_correction}
Incident and scattered photons in the sample undergo attenuation within the sample, an effected referred to as self-absorption. This absorption is due to their absorption with Auger emission and fluorescence~\cite{AmentRMP2011}. It depends on the photon's incident, scattered geometry, energy, and polarization. 
Therefore, to enable a meaningful comparison of the spectral weights obtained with different parameters, we performed this correction to the measured spectra.

The self-absorption correction protocol is adapted from Refs.~\onlinecite{Minola_PRL2015, Arpaia_ncom2023}. Considering the measured raw spectrum as $I(\omega_2)$ and the corrected real signal as $\eta(\omega_2)$, the self-absorption correction equation can be written as:
\begin{equation}\label{SA_corr1}
 I(\omega_2)  = \frac{\rho_{R1}}{\rho_{T1}} \cdot C(\alpha, \beta, \omega_1, \omega_2) \cdot \eta(\omega_2)
\end{equation}
where $\omega_2$ is the scattered photon energy, $\alpha$ and $\beta$ are the incident and scattering angles, as noted in Fig.~\ref{experiments_geometry} $C$ is the absorption coefficient and can be written as:
\begin{equation}\label{SA_coef}
C(\alpha, \beta, \omega_1, \omega_2) =  \frac{1}{1+ \frac{\rho_{T2}}{\rho_{T1}} \cdot \frac{sin(\alpha)}{sin(\beta)}}
\end{equation}
where $\rho_{T1}$ ($\rho_{T2}$) is the total X-ray absorption spectroscopy (XAS) intensity at incident (scattered) photon energy $\omega_1$ ($\omega_2$), $\rho_0$ is the XAS intensity on the pre-absorption-edge, and $\rho_{R1}$ is only the resonating part of the XAS, all noted in Fig.~\ref{XAS_absorption}.

Note that we only consider the ratio, not the absolute XAS intensity value in the absorption coefficient. Therefore, the prefactor is used as a normalization to account for the intensity variations between one XAS spectrum and another, for instance when varying the polarization.

Finally, the corrected spectrum can be extracted as:
\begin{equation}\label{SA_corr2}
\eta(\omega_2) = \frac{\rho_{T1}}{\rho_{R1}} \cdot (1+ \frac{\rho_{T2}}{\rho_{T1}} \cdot \frac{sin(\alpha)}{sin(\beta)} ) \cdot I(\omega_2) 
\end{equation}

The polarization dependencies are considered in the XAS where XAS has been measured using $\sigma$ ($\pi$) polarization to correct the measurement taken with this polarization.

\begin{figure*}[!htb]
    \centering
    \includegraphics[width=1\textwidth]{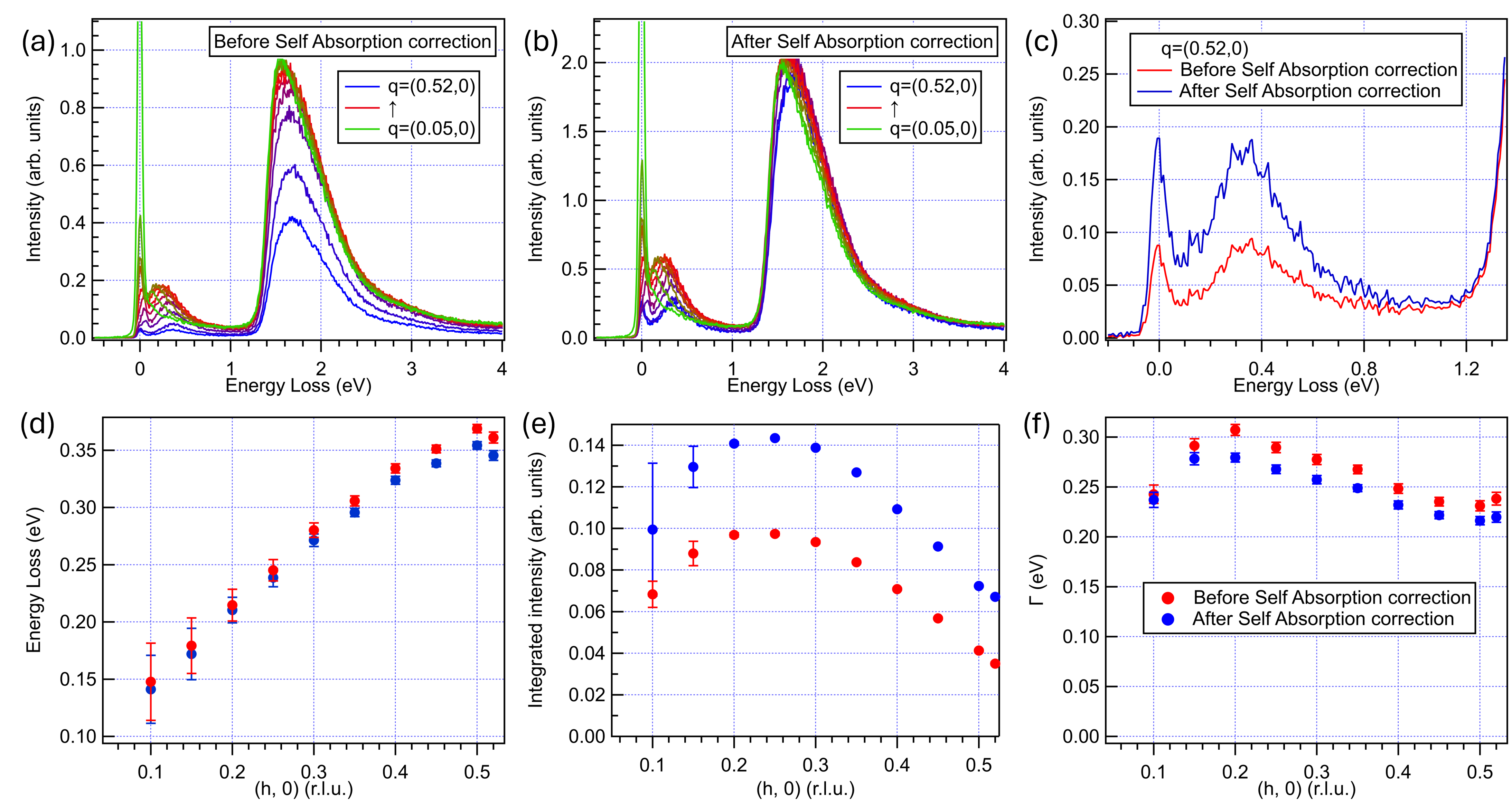}
    \caption{Self-absorption correction effect on RIXS spectra and fit parameters. See text for details.}
    \label{SA_correction}
\end{figure*}

The effect of the correction on paramagnon fit parameters is shown in Fig.~\ref{SA_correction}.  In Fig.~\ref{SA_correction}(a) and (b), we compare RIXS spectra taken for the paramagnon dispersion along the direction (h, 0) before (a) and after (b) the self-absorption correction, where we can see the effect on the full spectrum. The effect can be well seen in the \textit{dd-excitation}, where we no longer need to normalize their area. In Fig.~\ref{SA_correction}(c), we directly show the effect on the low-energy part of the spectrum, where only in this figure, we normalized to the \textit{dd} area to compare the Self-absorption correction. In (d)-(e)-(f), we show the effect of the correction on paramagnon fit parameters (Energy Loss $\omega$, integrated intensity, and width $\Gamma$ ). $\omega$ and $\Gamma$ are slightly different and, in some cases, may be considered the same within the error bar, but the integrated intensity changes drastically.

\begin{figure*}[!htb]
    \centering
    \includegraphics[width=0.45\textwidth]{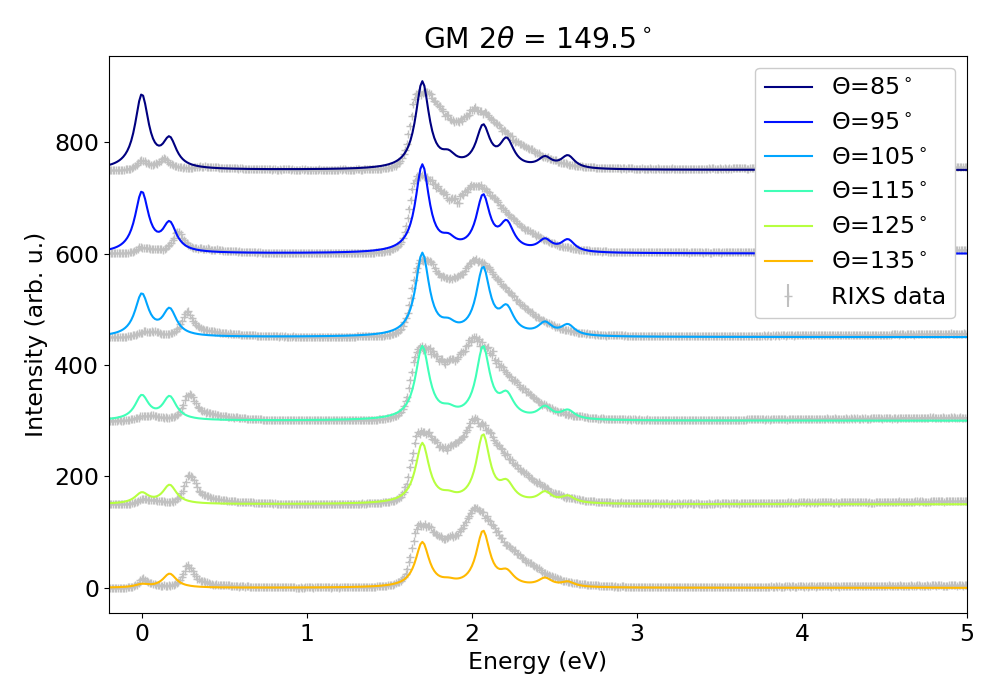}
    \includegraphics[width=0.45\textwidth]{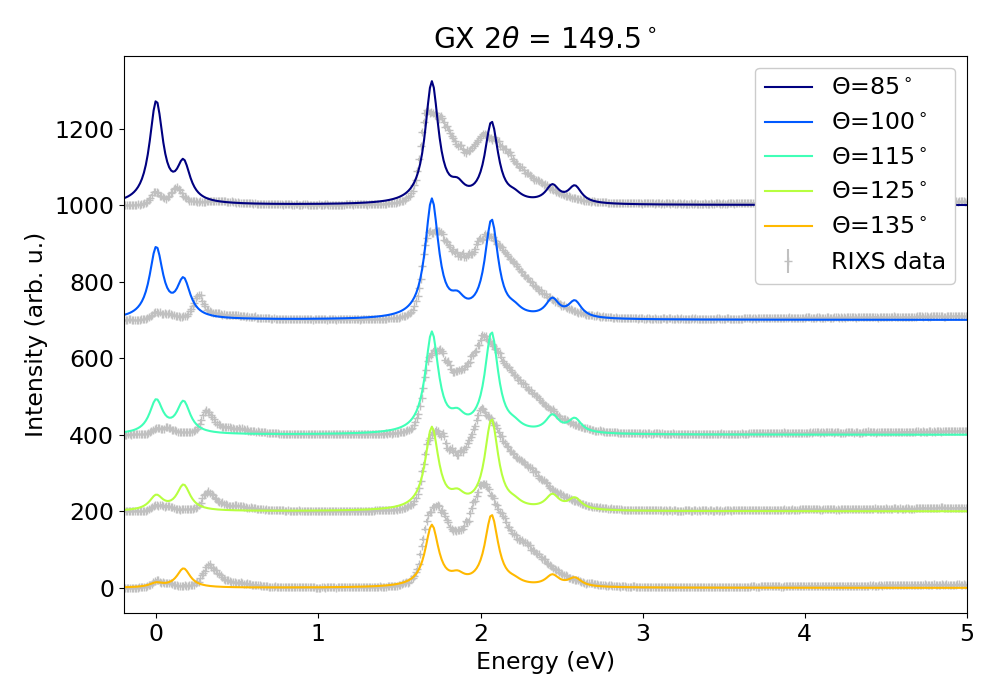}
    \caption{Crystal Electric Field multiplet calculation along the $\Gamma-M$ (left panel) and $\Gamma-X$ (right panel) directions, compared to RIXS measurements (raw experimental data from Ref. \onlinecite{Lebert_2023}), with self-absorption corrections described in Sec. \ref{Self_absorption_correction}.}
    \label{CEF_calc}
\end{figure*}

\subsection{Crystal Electric Field Simulations}

To validate the self-absorption corrections, as the intensity reference is the strong signal from crystal electric field multiplet \textit{dd} excitations, we performed a comparison between Crystal Electric Field (CEF) calculations and data in the undoped compound, taken from previous measurements reported in Ref. \onlinecite{Lebert_2023}, but normalized with the same protocol described in the previous Sec. \ref{Self_absorption_correction}. 

\begin{table}[ht]
\centering
\begin{tabular}{lcc}
\toprule
Parameter & initial/final state & Intermediate state \\
\midrule
$F_{dd}^2$ & $12.8555 \times 0.8$ & $13.612 \times 0.8$ \\
$F_{dd}^4$ & $7.981 \times 0.8$ & $8.4577 \times 0.8$ \\
$F_{pd}^2$ & --- & $8.1784 \times 0.8$ \\
$G_{pd}^1$ & --- & $6.1698 \times 0.8$ \\
$G_{pd}^3$ & --- & $3.5106 \times 0.8$ \\
$\xi_{3d}$ & 0.1019 & 0.1244 \\
$\xi_{2p}$ & --- & 13.4393 \\
$Dq$ & 0.17 & 0.17 \\
$Ds$ & 0.39 & 0.39 \\
$Dt$ & 0.16 & 0.16\\
$J_{ex}$ & 0.17 & 0.17\\
\bottomrule
\end{tabular}
\caption{Values of Quanty input parameters given in eV.}
\label{tab:slater}
\end{table}

The CEF model was implemented using the Quanty Package \cite{PhysRevB.90.085102,Haverkort_2014,PhysRevB.85.165113}. Assuming a weak impact of the doping on the structure, the Cu$^{2+}$ ions in Na-CCOC experience a $D_{4h}$ point group symmetry. In this setting, the local z axis lies along the $C_4$ axis, here corresponding to the crystallographic $\vec{c}$, and the remaining local axes $x$ and $y$ point along the crystallographic $\vec{a}$ and $\vec{b}$ axes, respectively. 
In the $D_{4h}$ symmetry, the crystal-electric field (CEF) splits the five 3\textit{d} orbitals into four energy levels denoted $a_{1g}$, $b_{1g}$, $b_{2g}$ and $e_{g}$, corresponding to the $z^2$, $x^2-y^2$, $xy$ and ($xz,yz$) orbitals, respectively. Their respective energies can be written as a function of three CEF parameters as follows:\\
\begin{align}
\begin{split}
    E_{a_{1g}} =& \phantom{-\ }6Dq - 2Ds - 6Dt\\
    E_{b_{1g}} =& \phantom{-\ }6Dq + 2Ds - \phantom{4}Dt\\
    E_{b_{2g}} =& -4Dq + 2Ds - \phantom{4}Dt\\
    E_{e_{g}}\  =& -4Dq - \phantom{2}Ds + 4Dt
\end{split}
\end{align}
In the case of Na-CCOC, the values $Dq~=~0.17$~eV, $Ds~=~0.39$~eV and $Dt~=~0.16$~eV were found to best reproduce the \textit{dd} excitations present in the spectra. This set of parameters generates CEF excitations at $1.70$~eV, $2.05$~eV and $2.35$~eV.
Covalency effects on the intra-atomic electron Coulomb interactions were accounted for by rescaling the 3\textit{d}-3\textit{d} and 2\textit{p}-3\textit{d} Slater integrals to $80~\%$ of their Hartree-Fock values. 
It is interesting to point out that the inclusion of spin-orbit coupling terms slightly lifts the degeneracy of the $e_g$ levels, leading finally to four excitations instead of the three expected from CEF alone. 
Finally, a Zeeman term was added to the single-ion model to mimic the antiferromagnetic exchange between Cu ions and thereby partially capture the magnetic interactions between them.
A Zeeman field amounting to an exchange coupling $J_{ex}$~=~0.17~eV between the $S_y$ spin components turned out to best reproduce the spectra. 
We note that this value is close, but does not match exactly the value obtained to reproduce the spin-wave dispersion in antiferromagnetic \CCOC\ \cite{Lebert_2023}, which includes contributions from higher exchange (or hopping) terms. 
Since the calculations are done in the single-ion approximation, they do not include dispersive spin-wave excitations.
However, the spin-flip excitation at $E\sim$~0.17~eV, roughly matches the average energy of the measured magnons.

The results are shown in Fig.~\ref{CEF_calc}. 
We nicely capture the inversion of the intensity ratio between the two main modes at 1.70~eV and 2.05~eV as a function of $\theta$. 
The overall intensity of the calculated \textit{dd} excitation decreases with increasing $\theta$, while this is not the case for the experimental data once corrected for self-absorption. 
However, the effect is much smaller than for uncorrected data (see Fig.~\ref{SA_correction}(a)), where a drop of about 60\% is observed.
The $\sim30-35\%$ drop in the calculated spectra is closer to the self-absorption-corrected data.

\subsection{Fitting procedure}\label{fit-description}
All RIXS data were fitted taking into account the following features in the spectra:
\begin{enumerate}[noitemsep, topsep=0pt]
  \item Elastic line: A Gaussian function
  \item Phonon: A Lorentzian function
  \item Spin-flip: A Damped Harmonic Oscillator (DHO) function
  \item Non-spin-flip: A Gaussian function
  \item \textit{dd}-excitation tail: A Gaussian function
  \item Background: A polynomial
\end{enumerate}
The widths of the elastic and the phonon lines was fixed to the resolution taken from the elastic line on a carbon tape.

\subsubsection{Damped Harmonic Oscillator function}
To fit the (para)magnon line shape, an anti-symmetrized Lorentzian was usually used. However, Lamsal \textit{et al.} \cite{Lamsal2016} brought to our attention that this function may be good in the case of undoped cuprates, where we have a sharp magnon, but can lead to inaccurate energy values in the case of doped cuprates, where the paramagnon is damped.
Therefore, we use a damped harmonic oscillator (DHO) here, which is more accurate than the anti-symmetrized Lorentzian and has been used more often recently~\cite{Lamsal2016, MonneyPRB2016,  PengPRB2018}.

It can be obtained from the imaginary part of the dynamical susceptibility \cite{PengPRB2018}, which describes the system's magnetic response at a frequency $\omega$. The DHO equation, corrected by the Bose factor, can be written as \cite{Li2024}: 
\begin{equation}\label{DHO}
\begin{split}
DHO(\omega) &=  \frac{A_0}{1-e^{- \omega/k_BT}} \cdot \frac{4 \omega \Gamma_q \cdot \omega_0}{(\omega^2-\omega_0^2)^2+(2 \omega \Gamma_q)^2} \\
& = \frac{A_0}{1-e^{- \omega/k_BT}} \cdot \frac{4 \omega \Gamma_q \cdot \sqrt{\omega_q^2 + \Gamma_q^2}}{(\omega^2-\omega_q^2-\Gamma_q^2)^2+(2 \omega \Gamma_q)^2}
\end{split}
\end{equation}
where $\omega_0$ and $\omega_q$ are the undamped and propagating magnon energy related by $\omega_q = \sqrt{\omega_0 - \Gamma_q}$, in which $\Gamma_q$ is the damping.

The Lorentzian function can be considered a limit of the DHO, as the latter is obtained by replacing $\omega_0$ in the Lorentzian with $\omega_q = \sqrt{\omega_0 - \Gamma_q}$. If 
$\Gamma_q$ is very small, as for an undamped magnon in undoped cuprates, we have $\omega_0 \approx \omega_q$. But in the overdamped regime of paramagnons, as in doped cuprates, we have $\omega_0 < \Gamma_q$, and the values of $\omega_0$ and $\omega_q$ diverge, leading to very different physical pictures.

\begin{figure*}[!htb]
   \centering
   \includegraphics[width=1\textwidth]{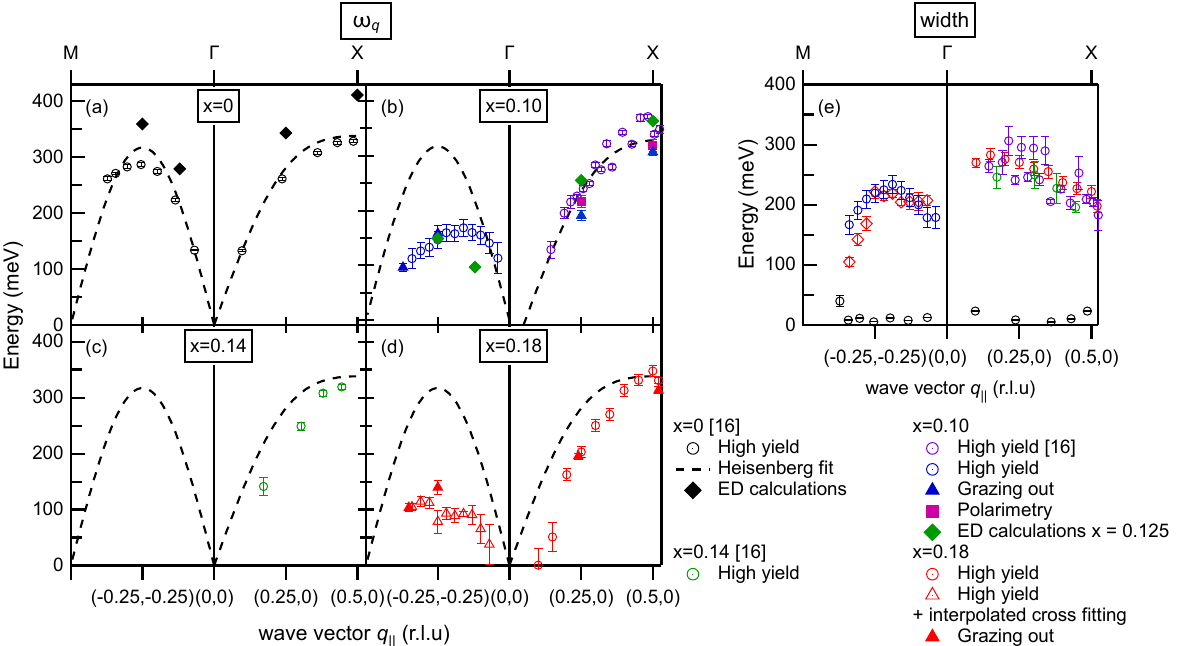}
   \caption{Doping dependence of the (para)magnon dispersion up to the maximum doping using $w_q$, including present and previously reported data \cite{Lebert_2023}. (a), (b), (c), and (d) correspond respectively to the $x=0$, $x=0.10$, $x=0.14$, and $x=0.18$ doping. Exact diagonalization (ED) calculations for $x=0$ and $x=0.125$ are shown in panels (a) and (b), respectively. The dashed line is the fitted Heisenberg model of the x=0 doping (AFM order) and is also plotted on the other dopings for comparison. (e) cDoping dependence of the (para)magnon width. Note that the data from \cite{Lebert_2023} were not corrected for self-absorption, unlike the data taken in this work. However, as shown in Fig.~\ref{SA_correction} in Appendix~\ref{Self_absorption_correction}, while this correction could affect the overall signal intensity, it has less impact on the peak lineshape, allowing us to reasonably compare the corresponding fitted parameters across datasets. Note also that the fit in Ref.~\onlinecite{Lebert_2023} was done using an anti-symmetrized Lorentzian, while in this work, we used a Damped Harmonic Oscillator.}
   \label{dispersion_all_wq}
\end{figure*}

However, fitting the RIXS intensity with this model for low paramagnon energies is problematic. 
In these cases, minimization tends to fall into local minima without physical meaning, where either $\omega_q$ or $\Gamma_q$ is fixed to $\sim 0$. Indeed, in such cases, the values of $\omega_q$ and $\Gamma_q$ can be interchanged, leading to the same lineshape of the fit. 
Therefore, we use $\omega_{max}$ corresponding to the energy at the maximum of the DHO lineshape, given by Ref.~\onlinecite{XiaotingLi_PRL2024} as:
\begin{equation}
    \omega_{max} = \sqrt{\frac{1}{3} \left(\omega^2_q - \Gamma^2_q  +2 \sqrt{\omega^4_q + \omega^2_q \Gamma^2_q + \Gamma^4_q} \right)  }
\end{equation}
and provides a more direct quantity for comparison with the model calculations. For completeness, we also show the corresponding fitted values of $\omega_q$ in Fig.~\ref{dispersion_all_wq}, which reproduces Fig.~\ref{dispersion_all} using $\omega_q$ instead of $\omega_{max}$.

\subsection{Cross fitting method}\label{Cross_fitting}
\begin{figure}[!htb]
    \centering
    \includegraphics[width=0.35\textwidth]{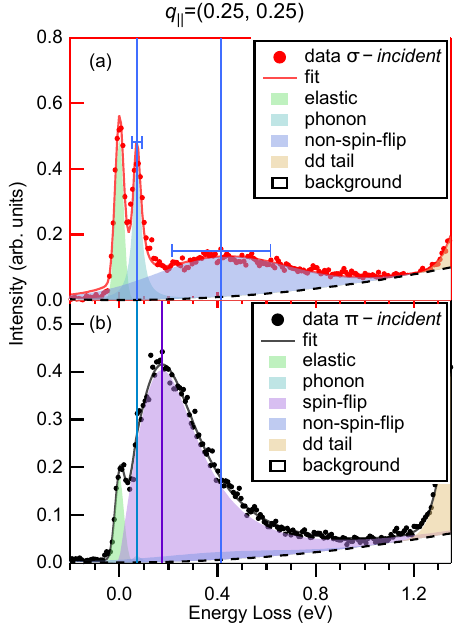}
    \caption{Cross-fitting. (a) $\sigma-incident$ spectrum fit. (b) $\pi-incident$ spectrum fit. The light blue vertical lines denote the positions of the phonon and multimagnon extracted from the $\sigma-incident$ data fit and fixed in the $\pi-incident$ fit, while the magenta ones indicate the fitted position of the para-magnon in the  $\pi-incident$ spectrum. The horizontal, capped bars indicate the width of the phonon and multimagnon extracted from the $\sigma-incident$ data fit and fixed in the $\pi-incident$ fit.}
    \label{cross_fit_ex}
\end{figure}

This method is used in the grazing-out geometry measurements. From the $\sigma-incident$ spectrum (Fig.~\ref{cross_fit_ex}a), where we have a dominant non-spin-flip signal, we extract the fit parameters for phonon and non-spin-flip contributions. Then, we fix these excitations' width and energy position in the $\pi-incident$ fit (Fig.~\ref{cross_fit_ex}b), leaving only their intensity as free fit parameters to separate them from the dominant spin-flip contribution. Since the single-ion model is used, the intensity ratio could not be quantified, and fixing a ratio between them would be a very bold assumption.

\section{Standard dispersion data}\label{Standard_disp}
\subsection{Paramagnon dispersion measurements for x=0.10}

Dispersion measurements were performed with the \textbf{high yield} method for RIXS. 
In such a configuration, only the highest q-vector happens to be in a \textbf{grazing-out} configuration. 
In that case, we can apply the \textbf{grazing-out} analysis with a cross-fitting of the contributions from the two polarizations. 

\begin{figure*}[!htb]
   \centering
   \includegraphics[width=1\textwidth]{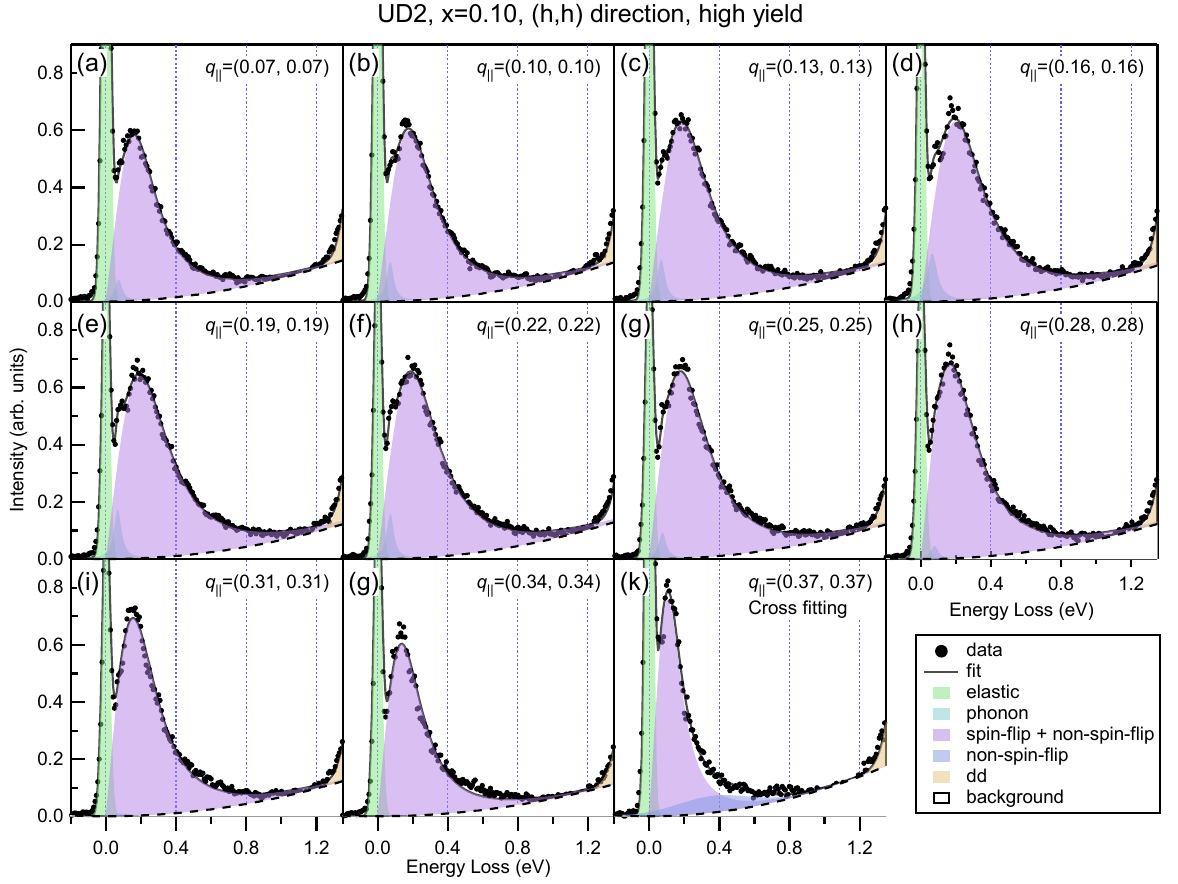}
   \caption{RIXS data and fits along $(h,h)$ direction for x=0.10 doping. The data, fit, background, and individual components correspond to the black dots, continuous black lines, the dashed lines, and the shaded areas on top of the background.}
   \label{GM_fit_10}
\end{figure*}

All the measurements and their fits are shown in Fig.~\ref{GM_fit_10}.

Since the \NSF\ signal appears weak at this doping, it was not considered for fitting procedures. One contribution in the magnetic excitation region, the paramagnon, was large enough to obtain a good fit except for high q-vectors (Fig.~\ref{GM_fit_10}i,g), where one can nevertheless see a shoulder between 400 and 800 meV.

\subsection{Paramagnon dispersion measurements for x=0.18}
\begin{figure*}[!htb]
   \centering
   \includegraphics[width=1\textwidth]{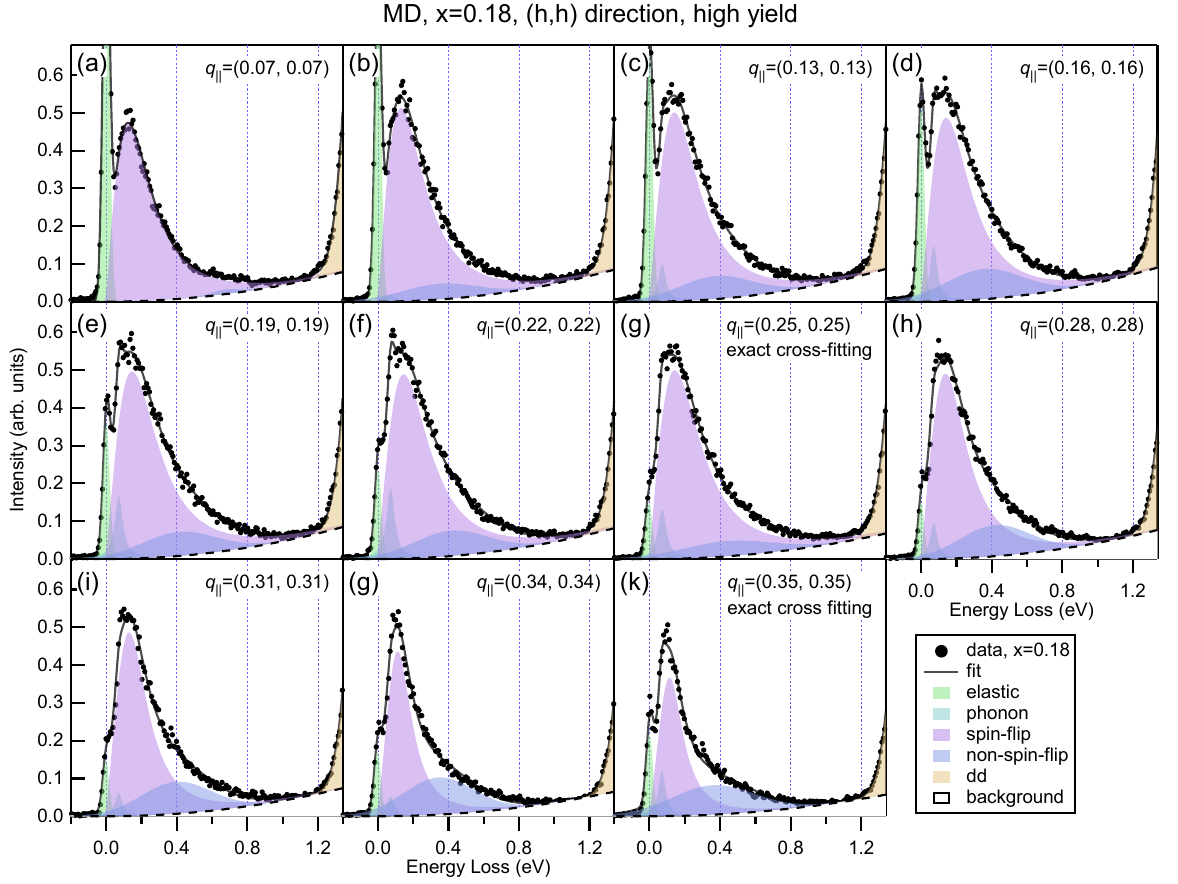}
   \caption{RIXS data and fits along $(h,h)$ direction for x=0.18 doping. The data, fit, background, and individual components correspond to the black dots, continuous black lines, the dashed lines, and the shaded areas on top of the background.}
   \label{GM_fit_18}
\end{figure*}

Measurements along the $(h,h)$ direction, using the \textbf{high yield} method, are presented in Fig.~\ref{GM_fit_18}. 
The presence of the \NSF\ signal in the \textbf{grazing out} measurements served as a way to separate this contribution from the \SF\ contribution, giving a high and low limit of its energy range. 
Therefore, the fit of the $(h,h)$ dispersion took into account two contributions: \SF\ and \NSF\ channels. 
This was possible due to the paramagnon's significant energy softening, pushing it away from the energy region of the non-spin-flip signal. 
We also note that the fit is not satisfactory without considering both contributions.

\begin{figure*}[!htb]
   \centering
   \includegraphics[width=1\textwidth]{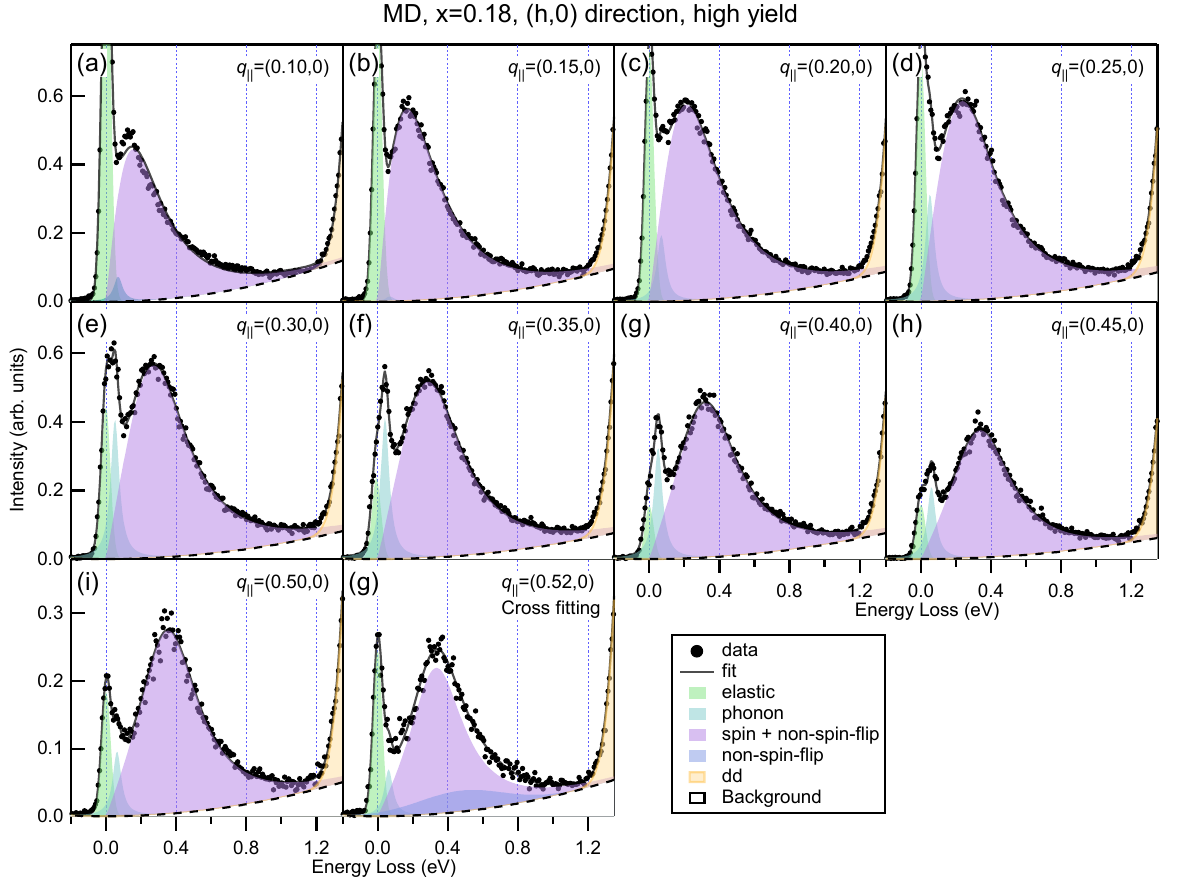}
   \caption{RIXS data and fits along $(h,0)$ direction for x=0.18 doping. The data, fit, background, and individual components correspond to the black dots, continuous black lines, the dashed lines, and the shaded areas on top of the background.}
   \label{GX_fit_18}
\end{figure*}

Along the direction $(h,0)$, Fig.~\ref{GX_fit_18}, the broad character of the paramagnon makes it harder to interpolate the energy position of the \NSF\ contribution; thus, the data were fitted only taking into account one feature, although we cannot rule out a weak \NSF\ contribution underneath the stronger \SF\ signal.

\section{\textbf{Grazing-out} method} \label{app:grazing_out_meth}
The dominance of the paramagnon signal using $\pi$-incident polarization in \textbf{grazing out} geometry was pointed out by Le Tacon \textit{et al.} \cite{LeTacon_nphys2011} as well as others by comparing experimental measurements and theoretical calculations \cite{Peng2015}.\\
Minola \textit{et al.} \cite{Minola_PRL2015}, used \textbf{polarimetry} measurements to demonstrate that in a \textbf {grazing-out} geometry, the $\pi$-incident polarization signal contains a dominant \SF\ process with a bit of \NSF\ contribution, and that the $\sigma$-incident polarization signal contains a dominant \NSF\ contribution. 
This method was used in several other studies, too, such as in Refs.~\onlinecite{Dean_nmaterials2013, Dellea2017, PengPRB2018}.

Li \textit{et al.} \cite{Li2024} also used this \textbf{grazing out} geometry method, but with the addition of a cross-fitting method to separate the \SF\ signal from the \NSF\ signal as explained in \autoref{Cross_fitting}.

In this work, we investigated all the above methods combined.
On the one hand, the \textbf{polarimetry} measurement confirms what was seen by Minola \textit{et al.} \cite{Minola_PRL2015}.
On the other hand, in \textbf{grazing-out} geometry, the cross-fitting method from Li \textit{et al.} \cite{XiaotingLi_PRL2024} gives a more appropriate separation of the \SF\ from the \NSF\ contribution. The reason for this is that even with a $\pi$-incident polarization, we might have a non-negligible contribution of the \NSF\ signal, as seen in our data (Fig.~\ref{Grazing_10} and Fig.~\ref{Grazing_18}), as well as in the work of Li \textit{et al.} \cite{XiaotingLi_PRL2024} and Minola \textit{et al.} \cite{Minola_PRL2015} for doped samples.\\
That being the case, the single-ion model is unable to account for the \NSF\ signal that appears in the \SF\ channel. In Table~\ref{main:pol_table2}, the \NSF\ should have a maximum intensity corresponding to $5\%$ of the \SF\ signal for an upper limit of 10$^\circ$ of the grazing angle. Meanwhile, we see much stronger contributions, as noted in Ref.~\onlinecite{Minola_PRL2015}.

Our analysis using the cross-fitting method shows good agreement with the \textbf{polarimetry} data within the error bars, as shown in Fig.~\ref{dispersion_all}. 
In conclusion, using \textbf{polarimetry}, we validate the method used in Ref.~\onlinecite{XiaotingLi_PRL2024} as being complementary to the grazing study done in Ref.~\onlinecite{Minola_PRL2015}. 

\section{Computational details}\label{Comput_det}

We compute the dynamical spin structure factor (DSSF) $S(\mathbf{q}, \omega)$ of the Hubbard model using exact diagonalization. 
At half filling, the Hilbert space dimension grows as $\binom{N_{\mathrm{orb}}}{N_{\mathrm{orb}}/2}^2$ with the number of orbitals $N_{orb}$.
This factorial growth of the Hilbert space restricts us to clusters of 16 orbitals to comfortably run calculations.
On finite clusters with periodic boundary conditions, the allowed momenta form a sparse discrete set of points in the first Brillouin zone determined by the geometry of the cluster.
This sparseness introduces two main drawbacks when computing observables on these clusters: (i) finite-size artifacts can introduce unphysical features in the computed spectra and (ii) the information outside of the discrete momentum points allowed by symmetry is unattainable. 
The first issue can be addressed by using twist-averaged boundary conditions, which were originally applied to Quantum Monte Carlo calculations~\cite{linTwistaveragedBoundaryConditions2001} and have been shown to successfully mitigate finite-size effects as well for exact diagonalization calculations~\cite{tohyamaAsymmetryElectronicStates2004}. 
To access other momentum transfer vectors and mitigate the second issue, we use a combination of different cluster shapes, each having its own set of allowed discrete momentum vectors. Since some of the clusters share common momentum transfer vectors, we can study the transferability of the results on different clusters.

\subsection{Twisted boundary conditions}

Twisted boundary conditions are parametrized by the twist vector $\boldsymbol{\kappa} = (\kappa_x, \kappa_y) \in [-\pi, \pi)^2$ which modifies the boundaries of the cluster in such a way that the wave function picks up a phase upon a translation by a lattice vector $\mathbf{a}_\mu$ : 
\begin{equation}
    \Psi(\mathbf{r} + \mathbf{a}_\mu) = e^{i\boldsymbol{\kappa}_\mu}\Psi(\mathbf{r})
\end{equation}

For a Hubbard Hamiltonian : 
\begin{equation}
    \hat{H} = -t \sum_{\langle i,j \rangle, \sigma} \cre_{i,\sigma} \anh_{j,\sigma} + U \sum_i \hat{n}_{i\uparrow}\hat{n}_{i\downarrow}
\end{equation}
The twist is expressed through a Peierls-like substitution $t_{ij} \to e^{i \mathrm{\kappa} \cdot \mathrm{r}_{ij}}t_{ij}$ (which is equivalent to the site-dependent gauge transformation presented in the main text) applied to all hopping terms, where $r_{ij}$ is the bond vector.

\subsection{Dynamical spin structure factor}

The longitudinal dynamical spin structure factor is defined as : 
\begin{equation}
    S^{z}(\mathbf{q}, \omega) = \sum_m \left| \left\langle \Psi_n \left| \hat{S}^z_\mathbf{q} \right| \Psi_0 \right\rangle \right|^2 \delta(\omega -E_m + E_0)
    \label{eq:dssf}
\end{equation}
where $S^z_\mathbf{q} = N^{-1/2} \sum_j e^{i\mathbf{q}\cdot\mathbf{r}_j}S^z_j$, $\left| \Psi_0\right\rangle$ ($\left| \Psi_m \right\rangle$) is the ground (excited) state and $E_0$ ($E_m$) its associated energy. 
The overlaps and energies of \autoref{eq:dssf} are computed using a simple Lanczos algorithm.

In practice, we replace the delta function by a Lorentzian with broadening $\eta$ 

\begin{equation}
    \delta(\omega) \longrightarrow \frac{1}{\pi} \frac{\eta}{\omega^2 + \eta^2},
\end{equation}
which is exact in the limit $\eta\rightarrow0^+$ and constitutes an approximation for finite values of the broadening $\eta$.

In order to mitigate part of the finite-size effects inherent to the cluster size, following the procedure outlined in Ref.~\cite{tohyamaAsymmetryElectronicStates2004}, we computed the DSSF for a set of twist vectors defined on a 16x16 Monkhorst-Pack grid (which was reduced to 16 unique calculations due to the symmetries of the clusters) and averaged the spectra.

\subsection{Raman signal}\label{Raman_signal_simulation}

The Raman scattering signal was computed in the $B_{1g}$ channel using the non-resonant response function \cite{jia2014a} for the Hubbard model: 

\begin{align}
    I_{B_{1g}}(\omega) &= - \frac{1}{\pi} \left\langle \Psi_0 \left| \gamma_{B_{1g}} \frac{1}{\hat{H} - E_0 - \omega + i 0^+} \gamma_{B_{1g}} \right| \Psi_0 \right\rangle \\
    \gamma_{B_{1g}} &= \sum_\mathbf{k} \left( \frac{\partial^2 \varepsilon(\mathbf{k})}{\partial k_x^2} - \frac{\partial^2 \varepsilon(\mathbf{k})}{\partial k_y^2}\right) \hat{c}_\mathbf{k}^\dagger \hat{c}_\mathbf{k}^{\phantom{\dagger}}\ ,
\end{align}
where $\left|\Psi_0\right\rangle$ is the ground state, $E_0$ its associated energy and $\varepsilon(\mathbf{k})$ is the non-interacting dispersion.

In the calculation of the Raman response, we only employ the 16q0 Betts cluster.
Indeed, as described in Table~\ref{tab:bets}, for all other clusters, the periodic boundary conditions introduce artifacts through the second-nearest neighbor hopping terms, which mix the $x$ and $y$ directions. 
As the Raman response depends on the derivatives of the dispersion, this will lead to unphysical results for all clusters except for the 16q0 cluster, for which no mixing between the $x$ and $y$ directions appears.

\subsection{Influence of the next-nearest neighbor hopping}

\label{sec:3NNhop}

\begin{figure*}
    \centering
    \includegraphics[width=0.95\linewidth]{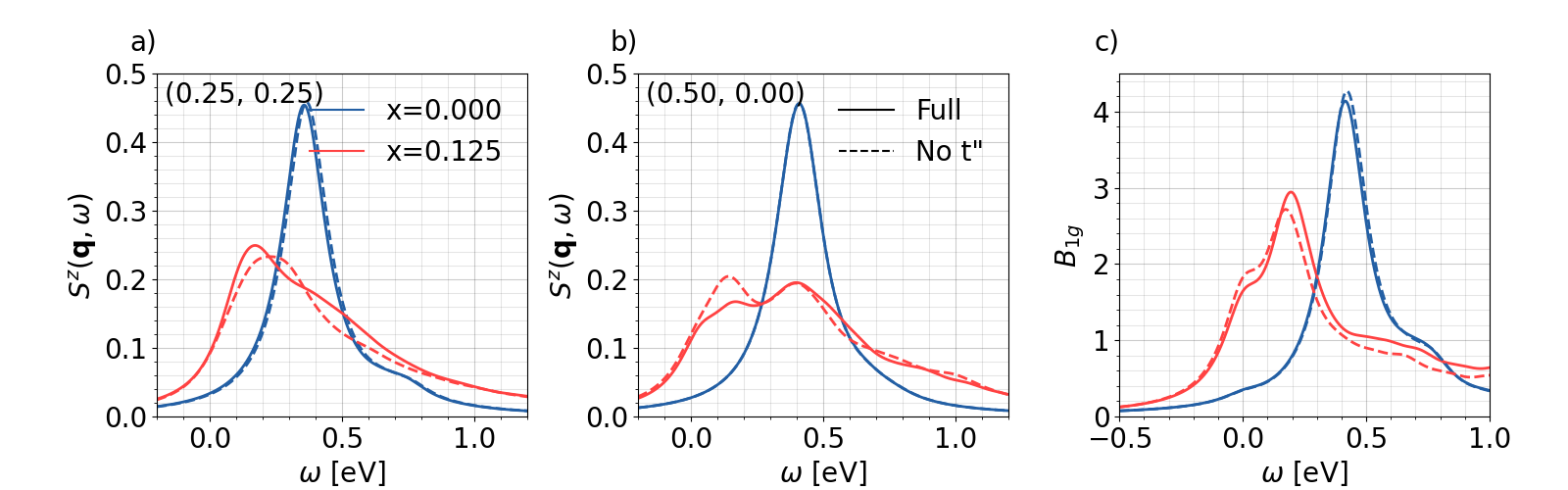}
    \caption{Dynamical spin structure factor (a and b) and Raman $B_{1g}$ intensity of the 16q0 cluster computed at two different doping values. In the absence of next-nearest neighbor hopping (dashed line), the softening along the paramagnon peak at q=(0.25, 0.25) is reduced while the peak position is unaffected at q=(0.5, 0.0). On the contrary, doping increases the softening of the multimagnon peak in the Raman signal.}
    \label{fig:tpp_softening}
\end{figure*}

In Fig.~\ref{fig:tpp_softening} we compare the dynamical spin structure factor of the full model (\autoref{eq:hamiltonian}, full lines) with the same model where $t^{\prime\prime}$ is explicitly set to $0$~eV (dashed lines). 
The absence of next-nearest neighbor hopping corresponds to the Hamiltonian chosen in a previous study \cite{jia2014a}.
At half filling (x=0, blue lines), we do not observe any notable differences in the dynamical spin structure factor at both $q=(0.25, 0.25)$ and $q=(0.5, 0.0)$. 
However, with doping, the softening of the paramagnon peak is significantly reduced (going from $\sim$ 0.2 eV to $\sim$ 0.1 eV) at $q=(0.25, 0.25)$, while the peak position is unaffected at $q=(0.5, 0.0)$. 
Similarly, the next-nearest neighbor hopping appears to have an influence on the computed Raman spectrum upon doping.
However, contrary to the DSSF, the absence of this hopping increases the softening of the signal, see Fig.~\ref{fig:tpp_softening}(c). 
This finding is consistent with the results of Ref.~\cite{jia2014a}, where the next-nearest neighbor hopping was not included in the model, and for which no significant softening of the paramagnon peak was reported at $q=(0.25, 0.25)$. 
This result highlights the importance of the next-nearest neighbor hopping in the mediation of this process.

\subsection{Cluster-by-cluster comparison}

As a means to further mitigate finite-size effects and isolate physical features from cluster-specific artifacts, we also averaged the computed DSSF on different cluster shapes. We chose to this end six different Betts clusters, see Table~\ref{tab:bets}.

\begin{table}[ht]
    \centering
    \begin{tabular}{|c|c|c|c|c|c|}
    \hline
         Cluster & $\mathbf{a_1}$ & $\mathbf{a_2}$ & Néel & \multicolumn{2}{c|}{Artifacts} \\
         \hline
         16q0 & [4 0] & [0 4] & Yes & $(2,0) = (-2,0)$ & $(0,2) = (0,-2)$ \\
         16d4 & [4 2] & [0 -4]& Yes & $(2,0)=(2,-2)$ & $(0,2) = (0,-2)$\\
         16h4 & [4 1] & [0 -4]& No & $(2,0) = (-2,1)$ & $(0,2) = (0,-2)$ \\ 
         16h6 & [4 -2] & [-2 -3] & No & $(2,0) = (-2,2)$ & $(0, 2) = (-2,-1)$ \\
         16h7 & [2 -2] & [3 5] & Yes & $(2,0) = (0,2)$ & $(0,2) = (2,0)$\\ 
         16d2 & [4 4] & [2 -2] & Yes & $(2,0) = (0,2)$ & $(0,2)= (2,0)$\\
         \hline
    \end{tabular}
    \caption{Specification for the six different Betts clusters used in this article. The Néel column reflects whether or not a specific cluster can stabilize a perfect antiferromagnetic order. The last columns represent the finite-size artifacts introduced on the third-nearest neighbor hopping ($t^{\prime\prime}$) due to the cluster shape.}
    \label{tab:bets}
\end{table}

\begin{figure}
    \centering
    \includegraphics[width=0.95\linewidth]{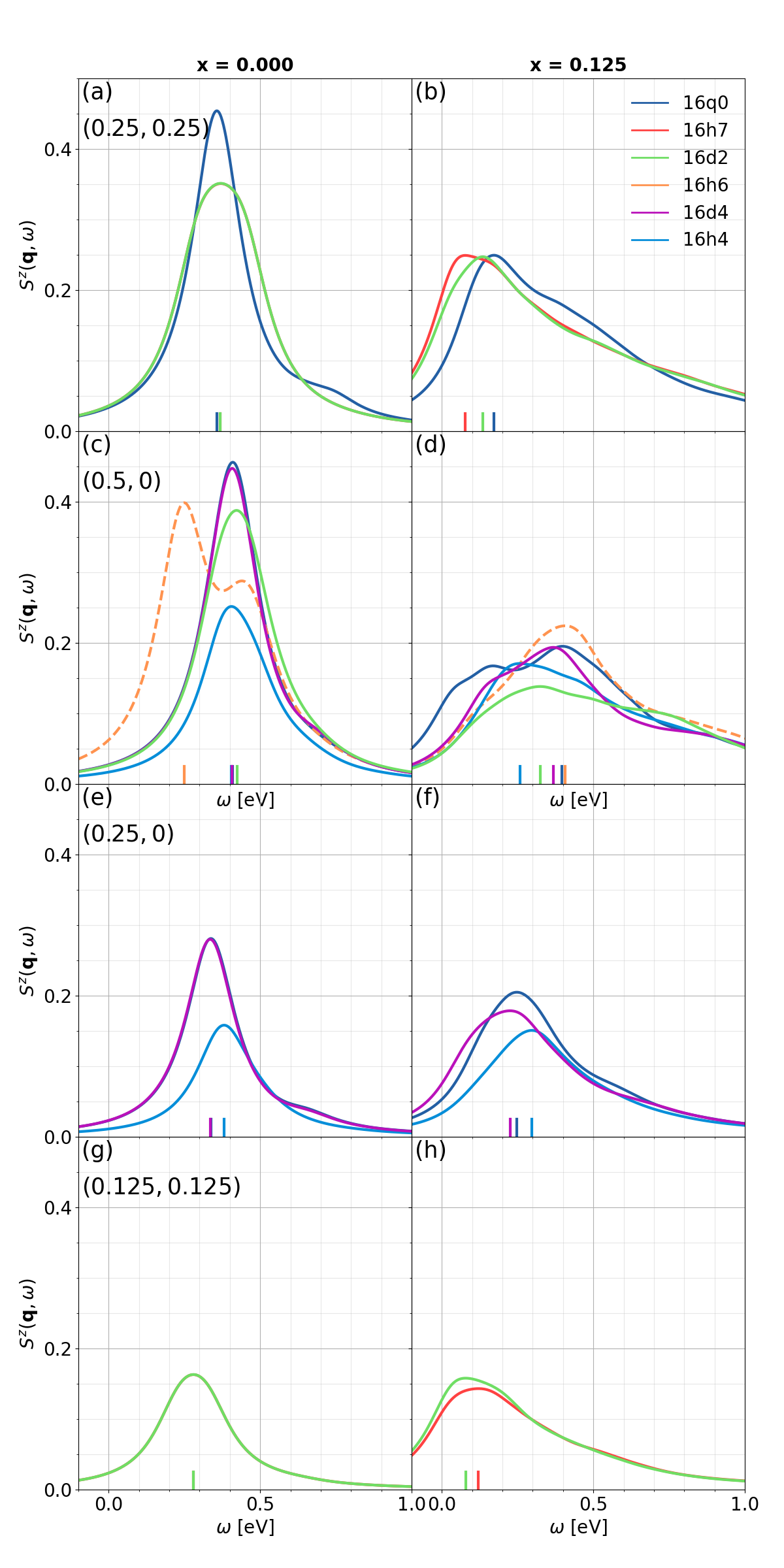}
    \caption {Cluster-by-cluster comparison of the dynamical spin structure factor for undoped (a,c,e,g) and 12.5\% doped (b,d,f,h) cases. Vertical lines on the $\omega$ axis represent the frequency of maximum intensity for each cluster.}
    \label{fig:clusterbycluster}
\end{figure}

In Fig.~\ref{fig:clusterbycluster}, we report the DSSF spectra obtained for the individual clusters (in colors) as well as the average of all clusters (black line).

For nearly all clusters, we observe a very good transferability for the different dopings and momentum transfer vectors. 
The notable exception is the 16h6 cluster (dashed orange line), which shows a peculiar double peak structure at the $X$ point in the undoped system. 
This double peak structure can be understood by looking at the geometry of the cluster: due to its particular lattice vectors, it can stabilize an antiferromagnetic order in only one direction, which would suggest two different possible magnetic spin excitations that appear in the spectrum in the form of two peaks.
As the other clusters do not exhibit this particularity, it allows us to rule out the relevance of this low-energy peak for our system.

\end{document}